# Multisensory Hybrid TV

Fernando Boronat, Lluc Simó, Ángel González, Pau Noguera
Universitat Politècnica de València – IIM R&D Group
Campus de Gandia
Calle Paraninfo, 1, 46730 Grao de Gandia (Spain)
fboronat@dcom.upv.es; lsimmar@upv.es; agondea@upv.es; pnogrod@upv.es

Corresponding author: Fernando Boronat (fboronat@dcom.upv.es)

**Abstract.** Traditional TV media content exclusively involves audiovisual streams. Multi-sensorial media (mulsemedia) introduces sensory effects such as ambient lighting, vibration, scents, wind, or temperature to stimulate other senses and create more enjoyable, realistic, and immersive content consumption experiences. This paper presents an MPEG-V- and HbbTV-based end-to-end solution for providing multi-sensorial hybrid (broadcast/broadband) TV content consumption experiences. A prototype following the proposal has been implemented on an end-to-end HbbTV-compliant platform, including both video streams and sensory effects metadata delivery and synchronous consumption, replicating a mulsemedia hybrid TV service. The prototype has been subjectively evaluated and validated by 106 participants with diverse profiles, yielding highly satisfactory results, including overall enjoyment, realism, immersion, perceived synchronization, and QoE enhancement. Furthermore, the results have also revealed interesting demographic patterns, such as participants with higher academic levels reporting lower satisfaction scores in general, distinct gender-based preferences regarding sensory acceptance, and potential barriers to commercial adoption, where higher realism and immersion do not translate into an equivalent willingness to pay for these enhanced TV services.

**INDEX TERMS:** Augmented TV, Digital TV, HbbTV, Hybrid TV, Immersive TV, Mulsemedia, Multi-sensorial media, Multisensory integration, Sensory Effects, TV.

# 1 Introduction

In the television (TV) context, broadband (i.e., IP-based) delivery is already having a real impact, offering viewers unrivaled choice and flexibility. However, despite the massive increase in content consumption over broadband networks, traditional broadcast TV remains a primary source of content. The less-noticed and less-discussed trend is that DVB-based Digital Terrestrial Television (DTT) continues to grow its overall number of TV households over the years and to evolve and perform very well in the market. DTT has served people well for so long and remains the mainstay of modern TV for many segments of viewers. The combination of DTT and IP into a hybrid (broadcast/broadband) environment is a model set to remain relevant for present-day consumers. The evolution to a Hybrid TV model has meant that DTT is now very well placed to underpin the market as it moves towards a pay-lite model. The content of linear broadcast and/or broadband TV can be enhanced with IP on-demand content. Viewers can receive a free base of content delivered via DTT and overlay IP-delivered offers, catch-up, and Subscription Video on Demand services from a range of providers.

The success and viewers' appetite for the hybrid model suggest that an IP-only future is still a very long way off. Instead, continuing to blend the best of IP and DTT into a workable hybrid offer that meets today's consumer needs is a model set to remain relevant for many years to come. For audiences, the result is a very reliable service built around a core broadcast service, with an enriched set of IP-delivered services. This offers additional content and flexible viewing options, including a more accessible, interactive, and increasingly personalized viewing experience. Accessibility services can be offered under this model to address the needs of people with hearing and low-vision impairments, learning difficulties, and older people. For example, adaptive subtitles in different languages, audio descriptions, and/or sign language videos in various formats can be provided as complementary content via IP networks. Moreover, users can personalize the playout of this kind of content (e.g., larger subtitles or placement near the speaking character).

The transition towards a connected or hybrid TV opens the door to new business models and opportunities. This is of great interest to TV operators and other interested parties (device manufacturers, content providers, etc.). However, despite the efforts of certain organizations, embodied in standards (e.g., HbbTV[1] [1] in Europe; ATSC[2] in America; or MMT[3] in Asia), and the old Smart TV Alliance[4], current solutions do not fully exploit the potential of hybrid TV services.

Regarding TV content, until now it has been viewed in 2D or 3D, with scenes and points of view selected by the production team. However, the introduction of new content types, such as omnidirectional (360) video, Augmented Reality (AR), and multisensory effects, into TV services will enable TV service providers to offer more interactive, immersive, and personalized hybrid TV viewing experiences. Consumers will be able to select the consumption devices, the type of media content (from a set of related content available for the TV service), and, for 360 video content, the field of view (FoV) for each scene. This will be possible thanks to the proliferation of consumer devices (smart TVs, tablets, smartphones, and AR or Virtual Reality -VR- head-mounted displays or HMDs) and their ability to connect to various access networks.

Traditionally, consumed media content has exclusively involved audiovisual (AV) streams, typically stimulating only the senses of sight and hearing, even though the world is perceived as having 5 senses. Nevertheless, we are witnessing the rapid development of new technologies that are expanding traditional AV media into multimodal applications. To generate more immersive media consumption experiences, new types of content, consumption devices, and solutions that stimulate senses beyond the traditional ones of sight and hearing can be used [2]. Multi-sensorial media (a.k.a. mulsemedia) content provides additional sensory effects during media consumption, aiming to give consumers greater realism and immersion. Some examples are smells (related to the olfactory and taste senses), wind effects, pressure, or vibrations (related to the touch sense). This technology can also provide greater social integration for people with specific needs, contribute to the creation of better educational programs, and open new avenues for commercial purposes, such as immersive advertising and premium entertainment [3]. When dealing with multisensory stimuli, the relationships among senses cannot be ignored due to cross-modal correspondences, a phenomenon in which features perceived by different senses are automatically associated, which affects human performance in perceptual discrimination and overall QoE. Mulsemedia has been a widely researched area for many years [3]. Early subjective evaluations investigated how scents affect users to assess whether this enhanced media influences the sensation of presence in VR scenarios. As consumption-related technologies evolve, new associated challenges must be faced, such as enhancing omnidirectional video content with additional sensory effects triggered when the user watches specific regions of the content. Recent studies in VR mulsemedia applications have investigated the existence of leading sensory cues (vision, hearing, and touch) in material recognition tasks, demonstrating that while vision often acts as the leading cue, the simultaneous presence of multiple synchronized senses yields significant advantages for immersion and realism. Therefore, it is very novel and interesting to provide complementary data on multisensory effects to be included in future TV consumption experiences. They help provide much more accessible, immersive, and realistic experiences.

---

[1] Hybrid broadcast broadband TV. https://www.hbbtv.org/ (last access: November 2026)
[2] The Broadcast Standards Association. https://www.atsc.org/ (last access: November 2026)
[3] MPEG Media Transport. https://www.mpeg.org/standards/MPEG-H/1/ (last access: November 2026)
[4] https://www.smarttv-alliance.org/ (last access: November 2026)

The worlds of immersive content and hybrid TV services can be united, maintaining compatibility with current standards. The broadband delivery will provide adapted content that will ensure accessibility in immersive TV viewing experiences. Figure 1 shows an example of a domestic multi-device hybrid TV consumption scenario, including 360 video, AR content, CGI (Computer-Generated Imagery) and multisensory effects. Hereafter, the term 'immersive content' will refer to all of them. The TV set receives, via broadcast, a video stream of a portion of the scene (selected by the production team), and additional content can be received via broadband (IP) networks. In the figure, the latter consists of 2 video streams of 2 areas of the scene (selected by users), displayed on a tablet and a laptop, and a 360 video stream of the complete scene, to be viewed in a VR HMD or a 360 video player. The user can watch the entire scene by turning his head, dragging his finger across the screen, or moving the mouse, respectively, and selecting a new FoV. CGI can provide additional information about omnidirectional content (e.g., arrows indicating the main FoV of the scene at each moment). The user can privately view TV-related content using an AR HMD (e.g., 3D models or personalized subtitles). Smell, vibration, wind, and cold/heat generators are included to improve immersion and the user's quality of experience (QoE).

There is still much research to be done in this field to ensure acceptable QoE levels and address many challenges. Examples are [4]: the adaptive distribution of broadband content, complementary content signaling, synchronization (abbreviated as sync, hereafter) of the playout of related media content, QoS (Quality of Service), QoE, scalability, creation of a sense of presence, etc. Among them, hybrid sync is one of the most important problems to be solved in the near future to achieve acceptance of Hybrid TV services [5]. The term hybrid sync is used when the media streams involved are distributed using different delivery technologies. It usually includes other types of media sync (see below) to ensure synchronized playout of content across one or more devices for consumption. Without proper sync mechanisms, the latency inherent in IP-based broadband networks can degrade perceived QoE when broadband content is complementary to content delivered via DVB technologies. The HbbTV standard [1] already provides mechanisms for harmonizing the delivery and consumption of broadcast and broadband TV-related content through connected TVs and companion devices [6].

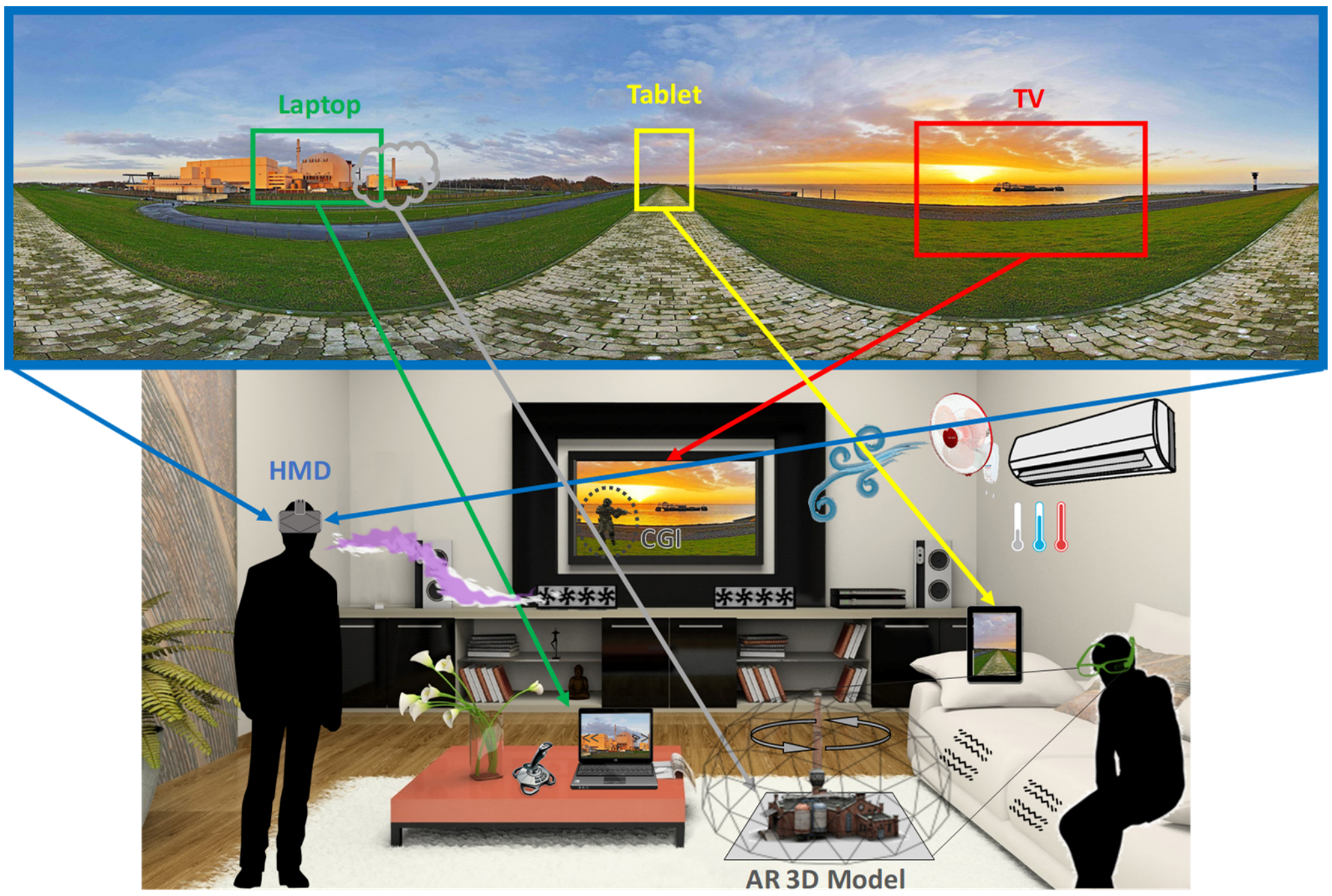

*Figure 1. Multi-device hybrid TV consumption scenario including immersive/accessible contents*

This paper focuses on one part of the scenario shown in Figure 1: the inclusion of synchronized multisensory effects in hybrid TV consumption experiences. An HbbTV-compliant approach to incorporating synchronized multisensory effects into TV services, a prototype implementation, and the results of a subjective user assessment are presented.

Figure 2 illustrates an example of the basic infrastructure required to establish a local mulsemedia environment. It consists of a media content (usually AV) player device (on the left side of the figure), with direct or indirect communication (via a controller device, in the center of the figure) with sensory effects generation devices (on the right side). The latter devices are responsible for generating different sensory effects around the user at the appropriate times, according to the viewed content (i.e., synchronized with the AV media playout) and considering environmental characteristics or conditions, which can be obtained from sensor devices. In multisensory hybrid TV systems, the main player device is typically a smart TV.

Previous works, described in the next section, have already presented end-to-end platforms to generate, deliver, and consume mulsemedia content, demonstrating its technical feasibility and evaluating basic QoE metrics. However, there is still a critical need to test the real viability of these systems in HbbTV-based scenarios with larger populations. Recent large-scale evaluations in 360-degree mulsemedia environments have demonstrated that demographic variables such as gender, age, prior computing experience, and sensory sensitivity significantly influence the user's perception [7]. For instance, differences in sensory sensitivity between genders reveal that females perceive scent and airflow effects more strongly than males. Furthermore, age plays an important role: younger users from Generation Z display a stronger engagement with multisensory content and tend to award higher scores to the mulsemedia experience. Prior media usage also influences QoE, as users who regularly watch dynamic video content rate aspects such as enjoyment and realism significantly higher than those who consume static content. While previous mulsemedia-related works successfully focused on technical sync, interactive authoring tools, and professional setups, the existing literature lacks comprehensive empirical data regarding actual home adoption in realistic broadcast environments [8].

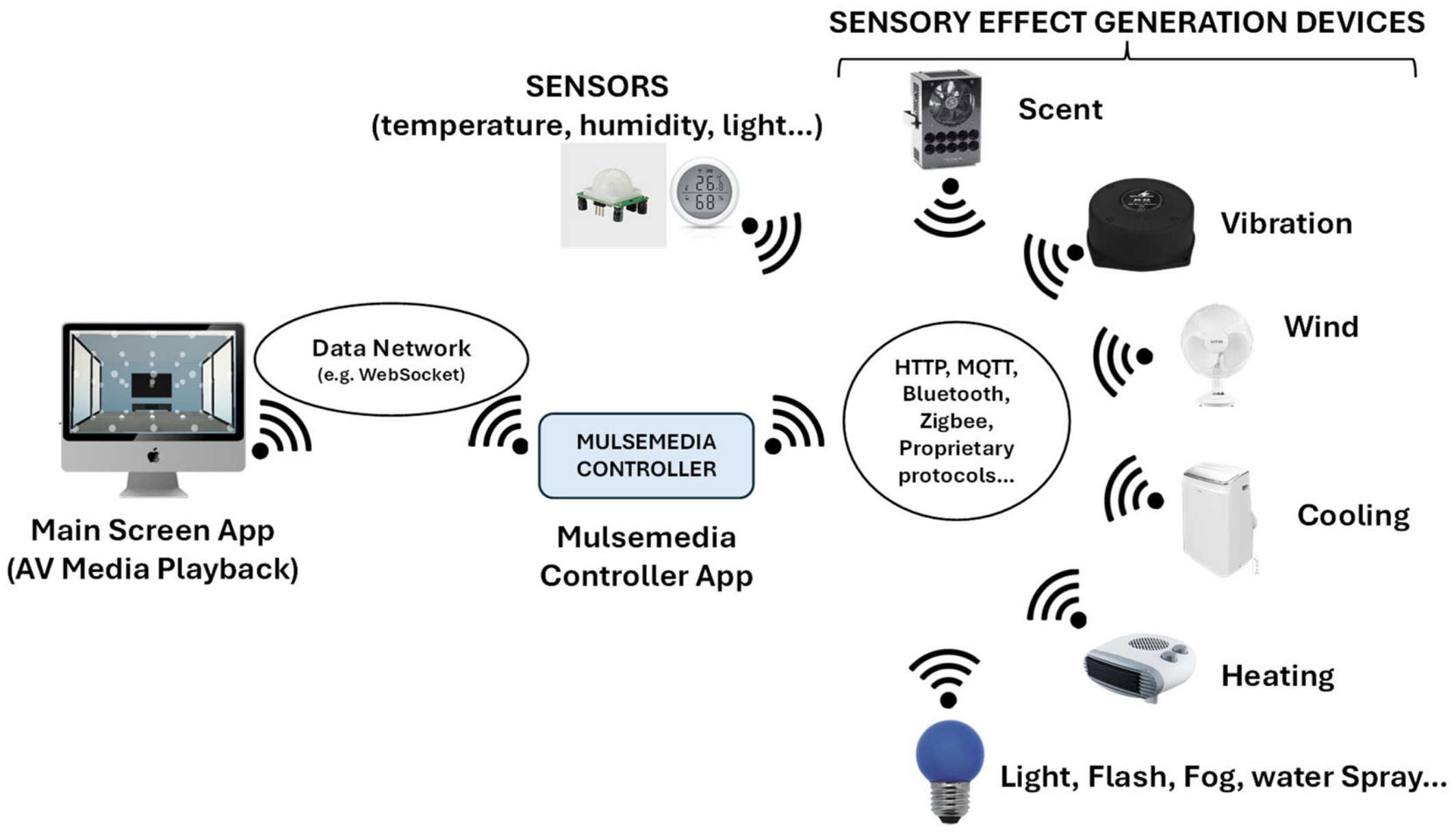

*Figure 2. Example of infrastructure required for a local mulsemedia environment.*

One of the main aims of this work is to conduct a subjective assessment of the practical viability of including mulsemedia in a typical domestic TV environment and to make an in-depth statistical analysis of the results obtained. By subjectively evaluating a real-world standard-compliant HbbTV-based mulsemedia TV scenario with a significantly large population (N=106), this work also analyzes the willingness to adopt this technology and which specific sensory effects are best accepted and tolerated by users, and how their responses vary based on demographic variables, such as gender, age, or educational level. To address these challenges, an end-to-end mulsemedia hybrid TV approach has been designed, and a prototype has been implemented by extending the authors' end-to-end HbbTV platform used in [8] to be MPEG-V[5] [9] and HbbTV standards-compliant and integrating the authors' mulsemedia platform described in [10].

As a summary, the main contributions of this work are the following:

- A complete MPEG-V-based and HbbTV-based end-to-end solution for providing multi-sensorial TV content consumption experiences.

[5] https://www.mpeg.org/standards/MPEG-V/ (last access: November 2026).

- Implementation of a prototype following the proposed solution. It includes content generation, the broadcast of DVB-T video streams, the broadband delivery of sensory effects metadata (SEM), and the synchronous consumption of TV content and generation of sensory effects at the consumer side. It uses an advanced multisensory testbed that provides several spatially distributed effects and emulates a home-like multisensory TV setting.
- Validation of an emulated multisensory hybrid TV service. A subjective evaluation (N=106 participants) was conducted to validate its correct performance and assess the users' acceptance and tolerance of sensory effects, enjoyment, immersion, distraction, fatigue, and QoE enhancement. An additional statistical analysis of the results has been conducted to identify behavioral patterns and detail how demographic variables, such as gender, age, and academic level, influence user satisfaction, opinions, and preferences.

The rest of this paper is organized as follows: Section II gives an overview of related work. The proposed solution is explained in Section III. The prototype implementation following it is described in Section IV. The evaluation, results, limitations, and a brief discussion are included in Section V. Finally, Section VI provides some concluding remarks and future work.

# 2 Related Works

To enable the creation of mulsemedia consumption experiences, it is first necessary to generate metadata describing the associated sensory effects, commonly referred to as Sensory Effect Metadata (SEM), and to establish an association between this metadata and the corresponding audiovisual (AV) content. Once linked, the AV content and its associated metadata can be distributed and consumed by mulsemedia systems capable of interpreting this information. Furthermore, communication among the different components of a mulsemedia system is required, and this communication must follow predefined formats and protocols. The MPEG-V: Media Context and Control ([9]) international standard (hereafter referred to as MPEG-V) specifies a format for representing this information, although it does not define procedures for exchanging it. Based on this standard, several authoring and simulation tools have been developed to help designers create and verify SEM Metadata.

Apart from MPEG-V, the Mulsemedia TV solution presented in this paper is based on other standards, such as DVB-CSS ([11] [12] [13]) and HbbTV ([1]). Moreover, in a multi-device scenario, a device discovery mechanism is needed. The DIAL protocol is used in the presented solution, as it is proposed for use in the HbbTV specification [1]. A brief overview of the mentioned standards, some tools for generating SEM, the DIAL protocol, and previous research work on mulsemedia in broadcast TV is provided in this section.

## 2.1 The MPEG-V standard.

To successfully deliver and reproduce sensory effects across heterogeneous devices and scenarios, standardization is crucial. The MPEG-V provides a comprehensive architecture and associated metadata formats for the interoperability between virtual worlds and real-world sensors and actuators. Among many other things, MPEG-V provides a wide range of tools for describing and representing sensory effects and devices. It provides tools for describing real and virtual worlds and specifies a series of binary structures and XML (eXtensible Markup Language)-based formats for exchanging data between the real world (the physical setting or installation where the multimedia application runs and is perceived by users) and the virtual world (the application itself). The standard consists of 7 parts. As Parts 2, 3, and 5 are directly related to the proposal in this article, they are briefly summarized in this section, including examples of the proposed syntax.

### 2.1.1 MPEG-V Part 2, ISO/IEC 23005-2 Control information [14]

This part defines the XML schema for representing the capabilities of effect generation and sensor devices (i.e., their properties), as well as the user's preferences regarding their characteristics and activation. It introduces the XML-based Control Information Description Language (CIDL), which provides the syntax for describing the fundamental structure of control information. To achieve this, several vocabularies are employed. The Device Capability Description Vocabulary (DCDV) defines the XML syntax for describing the capabilities of effect generation devices. Figure 3 presents an example description of a device capable of producing light in four colors (white, red, blue, and green), with a maximum illumination of 300 lux and 10 available intensity levels. The Sensor Capability Description Vocabulary (SCDV) supports describing sensor device capabilities, including light, humidity, temperature, and speed sensors (Figure 4). The Sensory Effect Preference Vocabulary (SEPV) provides a mechanism for expressing the user's preferences regarding particular sensory effects, such as lighting (Figure 5), wind, scent, and vibration. Finally, the Sensor Adaptation Preference Vocabulary (SAPV) allows the specification of the user's preferences for particular sensors based on the types of information they detect, as shown in Figure 6 for a light sensor.

```
<cidl:SensoryDeviceCapability
        xsi:type ="dcdv:LightCapabilityType"
        id="light1"
        unit="urn:mpeg:mpeg-v:01-CI-UnitTypeCS-NS:lux"
        maxIntensity="300"
        numOfLightLevels="10"
        locator="urn:mpeg:mpeg-v:01-SI-LocationCS-NS:right">

    <cdcdv:Color>
        urn:mpeg:mpeg-v:01-SI-ColorCS-NS:white
    </dcdv:Color>

    <cdcdv:Color>
        urn:mpeg:mpeg-v:01-SI-ColorCS-NS:red
    </dcdv:Color>

    <cdcdv:Color>
        urn:mpeg:mpeg-v:01-SI-ColorCS-NS:blue
    </dcdv:Color>

    <cdcdv:Color>
        urn:mpeg:mpeg-v:01-SI-ColorCS-NS:green
    </dcdv:Color>

</cidl:SensoryDeviceCapability>
```

*Figure 3. Example of the use of DCDV to describe the capabilities of a light effect generation device [14]*

```xml
<cidl:SensorDeviceCapability xsi:type="scdv:LightSensorCapabilityType"
        id="LS001" maxValue="400" minValue="0" numOfLevels="50" offset="1.5">

    <cidl:Accuracy xsi:type="cidl:ValueAccuracy" value="10.0"/>

    <scdv:Color>#FF000</scdv:Color>

    <scdv:Location>
        <mpegvct:X>1.0</mpegvct:X>
        <mpegvct:Y>1.0</mpegvct:Y>
        <mpegvct:Z>-1.0</mpegvct:Z>
    </scdv:Location>

</cidl:SensorDeviceCapability>
```

*Figure 4. Example of the use of SCDV to describe the capabilities of a light sensor device [14]*

```xml
<cidl:USPreference xsi:type="sepv:LightPrefType" activate="true"
        unit="urn:mpeg:mpeg-v:01-SI-UnitTypeCS-NS:lux" maxIntensity="300">

    <sepv:UnfavorableColor>
        urn:mpeg:mpeg-v:01-SI-ColorCS-NS:alice_blue
    </sepv:UnfavorableColor>

</cidl:USPreference>
```

*Figure 5. Example of the use of SEPV to describe the user's preferences for a light effect generation device [14]*

```xml
<cidl:SAPreference xsi:type="sapv:LightAdaptationPrefType" activate="true"
        id="LSAP001" maxValue= "400" minValue="10" numOfLevels="10"
        unit="urn:mpeg:mpeg-v:01-CI-UnitTypeCS-NS:lux"
        sensorAdaptationMode="scalable">

    <sapv:UnfavorableColor>#FF0000</sapv:UnfavorableColor>

    <sapv:UnfavorableColor>#FF0011</sapv:UnfavorableColor>

</cidl:SAPreference>
```

*Figure 6. Example of the use of SAPV for the adaptation of a light sensor device [14]*

### 2.1.2 MPEG-V Part 3, ISO/IEC 23005-3 Sensory information [15]

This part defines the representation of sensory effects to be produced by effect generation devices. It standardizes both the syntax and semantics of these effects, establishing the fundamental structures of sensory information required to generate SEM metadata. These structures are composed of reusable building blocks and shared attributes (e.g., effect type, duration, intensity, etc.) that are common to all defined effects. To represent these structures, the standard specifies the XML-based Sensory Effect Description Language (SEDL). Its attributes enable defining when a particular effect should be triggered, the intensity at which it should be generated, its duration, and other relevant parameters. Sensory effects are described using the Sensory Effect Vocabulary (SEV), which provides a detailed specification of each effect together with its general attributes (such as activation information, duration, priority, and position) and effect-specific attributes (e.g., color values for light effects or scent types for scent effects). Figure 7 illustrates the use of SEV to specify the activation and deactivation of a red light effect. SEDL and SEV are intended to be used jointly to create SEM metadata or descriptions that can be interpreted by the Media Processing Engine (MPE). The MPE may be implemented either as an embedded device or as a general-purpose computer capable of processing these descriptions and controlling effect generation devices that comply with the standard. In addition, the Interaction Interface Description Language (IIDL) is provided to describe both the commands exchanged with sensory effect generation devices and the information supplied by sensors.

```
<sedl:Effect xsi:type="sev:LightType" intensity-value="50.0"
        intensity-range="0.00001 32000.0" activate="true"
        color="#FF0000" si:pts="0"/>

...

<sedl:Effect xsi:type="sev:LightType" activate="false"
        color="#FF0000" si:pts="28"/>
```

*Figure 7. Example of the use of SEV to define the activation and deactivation of a red color light effect*

In the MPEG-V spatial model, each sensory effect includes a common location attribute that defines the region from which the user is expected to perceive it. The model places the user at the center of the reference system and organizes the surrounding space into three planes (*front*, *midway*, and *back*). Each plane is further divided into three height levels (bottom, middle, and top), and each height level contains five horizontal positions (*left*, *centerleft*, *center*, *centerright*, and *right*), as illustrated in Figure 8. This spatial arrangement results in a total of 45 distinct locations, each represented by a combination of three terms separated by the ':' character. For instance, assigning the position attribute the value *right:bottom:front* indicates that the corresponding effect will be activated or deactivated in the front plane, at the bottom level (floor), and on the right-hand side of the user's three-dimensional space.

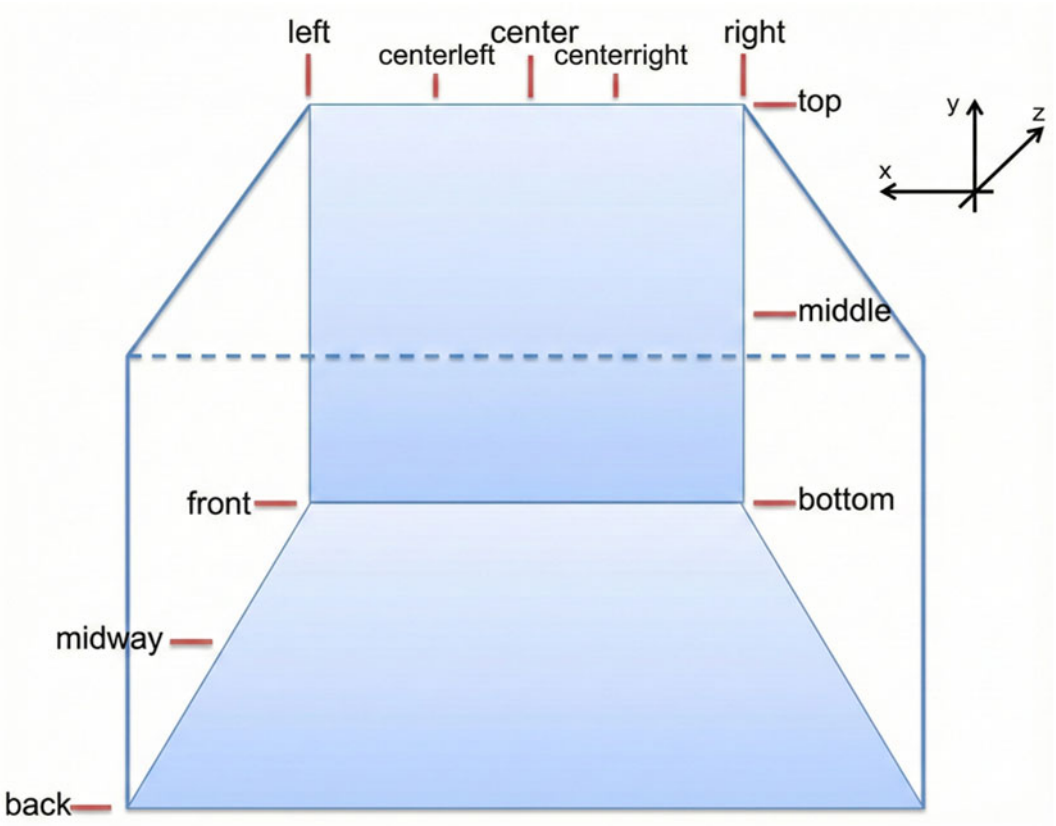

*Figure 8. MPEG-V spatial model for sensory effect metadata and reference coordinate system [16]*

### 2.1.3 MPEG-V Part 5, ISO/IEC 23005-5 Data formats for interaction devices [17]

This part defines the XML-based schema for representing both the information acquired from sensors (Sensed Information, SI) and the data types used in sensory effect activation and deactivation commands transmitted to effect generation devices. It standardizes the syntax and semantics of the data formats used by interaction devices, covering both commands for effect generation devices and SI. To support this functionality, it specifies the Interaction Interface Description Language (IIDL), which incorporates two additional vocabularies: the Device Command Vocabulary (DCV) and the Sensed Information Vocabulary (SIV). The Device Command Vocabulary (DCV) defines the messages used to control effect generation devices, including lights (Figure 9), fans, heaters, and scent generation devices. Whereas the SEV is intended to describe sensory effects using more elaborate structures (e.g., groups of effects), the DCV provides a simpler, more compact representation of sensory effects that can be readily interpreted by effect generation devices. The Sensed Information Vocabulary (SIV), in contrast, specifies the interfaces used to represent information generated by sensors, including light (Figure 10), temperature, and speed sensors.

```
<iidl:DeviceCommand xsi:type="dcv:LightType" id="light1"
        color="urn:mpeg:mpeg-v:01-SI-ColorCS-NS:red"
        intensity="5">

    <iidl:TimeStamp xsi:type="mpegvct:AbsoluteTimeType"
        absTime="1:30:23"/>

</iidl:DeviceCommand>
```

*Figure 9. Example of the use of DCV with a red light generation device*

```
<iidl:SensedInfo xsi:type="siv:LightSensorType" id="LS001"
        sensorIdRef="LSID001" activate="true" value="200"
        color="#FF0000">

    <iidl:TimeStamp xsi:type="mpegvct:ClockTickTimeType"
        timescale="1000" pts="600000"/>

</iidl:SensedInfo>
```

*Figure 10. Example of the use of SIV with a light sensor device*

## 2.2 Generation of sensory effect metadata (SEM)

Several tools can be used to generate SEM metadata for AV media content. Some examples are SEVino2 (*Sensory Effect Video Annotation tool* [18] [19], STEVE [20], RoSE (*Representation of Sensory Effects*) Studio [21] , SMURF (*Sensible Media aUthoRing Factory* [22]), H-Studio [23] and Real 4D Studio [24]. However, most of them are obsolete, non-MPEG-V compliant, or no longer publicly available. Except for STEVE 2.0, which is based on the event-based sensory effect sync paradigm, the rest of them are based on the timeline-based sensory effect sync paradigm and allow users to graphically define SEM metadata (in the case of H-Studio, only for sensory effects of movement and vibration) for AV media content and save it to files. Using these tools is very simple and intuitive. Their descriptions and how to use them are out of scope for this paper. In [25], a more detailed review and comparison of them, along with several proposals for representing sensory effects and their characteristics, is presented.

## 2.3 DVB-CSS: Digital Video Broadcasting- Companion Screens and Streams

ETSI TS 103 286 specification defines the Digital Video Broadcasting (DVB) Companion Screens and Streams (CSS) framework (DVB-CSS), specifying the architecture, protocols, and interfaces that enable companion devices (such as smartphones or tablets) to discover TV devices over the same home network, exchange content identification and metadata, and achieve synchronized media playout across multiple screens. It is composed of three parts, defined in [11], [12] and [13], respectively. They provide a complementary set of solutions that can be combined to enable companion devices to support new user experiences for consuming broadcast and Content on Demand (CoD) services.

Media content can refer to both the broadcasts made by an operator, e.g., a TV program transmitted to the audience, as well as extra content that a broadcaster makes available to accompany the program [11]. This extra content may be generic information, such as websites, or it may include audio or audiovisual enhancements, interactive content (e.g., games or quizzes), etc. The proposal presented in this paper includes additional content, such as metadata on sensory effects to be generated during main broadcast TV content consumption.

DVB-CSS includes protocols that enable a TV to indicate to companion devices what content it is presenting and its current timeline, thereby supporting reliable, accurate, and timely interactions over the home network [26]. It enables the development of compelling, interoperable implementations of CS products. It considers the use of the *MPEG-TS Timed External Media Information* (TEMI [27]), in combination with WebSocket, UPnP, and SSDP protocols [26] [28]. In particular, it is composed of different interfaces, building blocks and protocols to achieve different purposes, among others: Device Discovery and Association; Content identification and exchange of state information, which implies the usage of the UPnP, Simple Service Discovery Protocol (SSDP) protocol for discovering the Main Screen application (MS App); WallClock (WC) Sync protocol (called CSS-WC) providing the means to use a clock sync algorithm to establish a best effort approximation of the MS App's WC at the Companion Screen application (CS App); Timeline Sync (TS) protocol (called CSS-TS, for which the exchange of timestamp information between the involved (close-by) devices is needed; and Events triggering (e.g., detection and triggering of 'Do It Now events' in the broadcast stream). The CSS-TS protocol is used by the MS App to report on time positions on its timeline (timelines signaled in the broadcast or derived from the streaming container format).

## 2.4 HbbTV Standard

HbbTV ([1]) is an international initiative designed to unify the delivery of entertainment services over broadcast and broadband networks for connected televisions, set-top boxes, and companion devices. On the one hand, the main TV device (or set-top box) runs an application called the Main Screen Application (MS App), which manages the playout of the broadcast main content. On the other hand, Companion Screen applications (CS Apps) can run either on the same TV device (hybrid terminal) or on companion devices, managing the playout of the complementary broadband content. Since release 2.0.1, it has supported sync of both, on either a single device or across multiple devices. Within the scope of this work, its main features include:

- It relies on conventional DVB-T services (and therefore MPEG2 Transport Streams or MPEG2-TS) for the transmission of broadcast content.
- It adopts modern web technologies, including HTML5, CSS (Cascading Style Sheets), JavaScript, and WebSocket.

- It uses HTTP-based adaptive streaming (e.g., MPEG-DASH) as the broadband delivery technology for providing additional related media content.
- It adopts DIAL for device and service discovery. The discovery mechanism is based on the SSDP protocol, which is used to locate DIAL servers on the local network and access their DIAL REST (*REpresentational State Transfer*) services. It is also used to obtain the endpoint locations required for Application-to-Application communication (described below) and IDES.
- It provides an Application-to-Application (App2App) Communication Service based on a WebSocket Server (WSS), enabling direct communication between MS and CS Apps.
- It defines the use of an Application Information Table (AIT), embedded within the DVB MPEG2-TS, to signal the availability of HbbTV applications. The AIT is a critical component of HbbTV, providing essential metadata that enables the discovery, signaling, and management of interactive applications. It serves as a bridge between traditional broadcast services and broadband-delivered applications, informing HbbTV-compatible devices (such as smart TVs or set-top boxes) about the presence and behavior of applications that support both broadcast and broadband content. Content providers or broadcasters use the AIT to include URLs pointing to the applications that should be automatically launched when a user switches to a particular TV channel. When a user is watching a TV service (i.e., a channel), the AIT signals whether an associated HbbTV application is available. The AIT informs the receiver of the application's presence and requirements, and the TV can then automatically launch the application if it is capable of doing so.
- It supports (hybrid) sync between related media streams, or between media streams and applications, on a single hybrid terminal through the adoption of the TEMI solution [27].
- It also supports (hybrid) sync between related media streams, or between media streams and applications, across multiple devices in multi-screen environments by adopting the DVB-CSS specification described above.

In summary, HbbTV specifies certain functionalities for TV devices that are relevant to this work. It provides tools for TV service operators to enable smart TVs or Set-Top Boxes (STBs), i.e., MS devices, to run apps (referred to as HbbTV apps or MS Apps) that extend the broadcast linear content. The standard also provides tools for two-way communication between the MS App and the CS Apps running on different close companion devices. The MS device of the standard-compliant hybrid TV service is the smart TV/STB (referred to as a Hybrid Terminal), which can discover and be discovered by other close CS devices (running CS Apps). Once communication has been established, a CS App can receive information about complementary content related to the main broadcast TV content being consumed on the MS device, along with timing information from the MS App, which is useful for Inter-Device Media Sync (IDES) purposes.

## 2.5 DIAL protocol

The *DIscover And Launch* (DIAL) protocol [6] is the proposed protocol by the HbbTV specification to be used for device discovery. It provides a simple mechanism that allows companion devices (i.e., their CS App) to discover and launch applications on main devices (i.e., MS Apps). DIAL enables a seamless integration of CS Apps with the media consumption experience on the MS App. It relies on SSDP and HTTP protocols, and follows a client/server approach. It is composed of two parts: 1) DIAL Service Discovery, which enables a DIAL client to discover DIAL servers on its local home network and obtain access to the DIAL REST Service on those devices; and 2) DIAL REST Service, which enables a DIAL client to query, launch and, optionally, stop applications on a DIAL Server.

## 2.6 Mulsemedia in broadcast TV

Very few studies have been found related to the consumption of mulsemedia content delivered via DVB broadcast technology. Although they propose several ways to signal, embed, or deliver the sensorial media through DVB MPEG2-TS, this remains an additional challenge that must be addressed to provide the SEM Metadata associated with the broadcast content via this technology.

In [29] and [34], two different mulsemedia broadcasting systems, in which 3D and sensory information (in XML format) are encoded in MPEG-2 TS, are briefly explained (both are 2-page papers). The architecture, time-gathering methods, and time-sync methods for integrating 3D content with sensory information are described only very briefly and without detail. Similarly, in [30], the required end-to-end technology for broadcasting mulsemedia content based on the standard MPEG-V was proposed. The authors of the three mentioned works did not provide any implementation or evaluation of the proposals. In [31] and [32], a prototype of a smart home scenario including three sensory effects (cold, light, and vibration) was presented, implemented, and subjectively evaluated by 40 participants. Satisfactory QoE results were achieved, demonstrating the feasibility of multisensory TV services. The same researchers extended the subjective assessment, also with 40 participants, in [33], and instantiated and validated a parametric QoE model for multi-sensorial TV in a smart home scenario, thereby indicating the relationship between the quality of audiovisual content and user-perceived QoE for sensory effects applications. Unlike the approach presented in this paper, none of these works is based on the HbbTV standard.

In [8], the authors presented a preliminary, non-fully MPEG-V-standard-compliant basic solution for multisensory hybrid TV. It was implemented in an existing HbbTV-based end-to-end platform for hybrid TV content generation, delivery, and synchronous consumption. As an HbbTV terminal, a Linux-based emulator running on a computer was used, since no commercial HbbTV v2.0.1 terminals were available at the time. That setup demonstrated the technical feasibility of delivering MPEG-V compliant SEM metadata and enabling ad-hoc Inter-Device Media Sync (IDES) between the main broadcast TV content playout and the generation of associated sensory effects in HbbTV-compliant systems. The use of a dedicated gateway, called Mulsemedia Controller, was proposed. The approach proposed in this article is an extended and enhanced version of that preliminary proposal and already fully complies with the MPEG-V and HbbTV standards.

# 3 Proposed Solution for Multisensory Hybrid TV

In this section, the proposed solution for providing an end-to-end MPEG-V- and HbbTV-based multisensory TV service is presented in detail. First, a general overview of the solution is provided, describing its different parts, the applications or modules they include, the devices used, and the main processes involved.

## 3.1 General overview

As mentioned above, the proposed solution in this paper is an evolution of the authors' previous work in [8]. It has been significantly upgraded in both hardware and software parts. The processes involved in mulsemedia generation, the signaling of MPEG-V standard-compliant sensory effects metadata related to the main broadcast TV content, following the HbbTV specification ([1]), and the functionalities required to enable the satisfactory integration of sensory effects in TV consumption experiences are described in this section. Specifically, the proposed approach comprises two parts: the *Content Provider* and the *End Consumer* sides. The Content Provider side includes content generation and transmission, as well as the provision of necessary companion services. The End Consumer side comprises the main broadcast TV content played on an HbbTV-compliant TV Device, as well as sensory effect management and synchronous generation around the consumer. Figure 11 shows an overview of the proposed end-to-end solution.

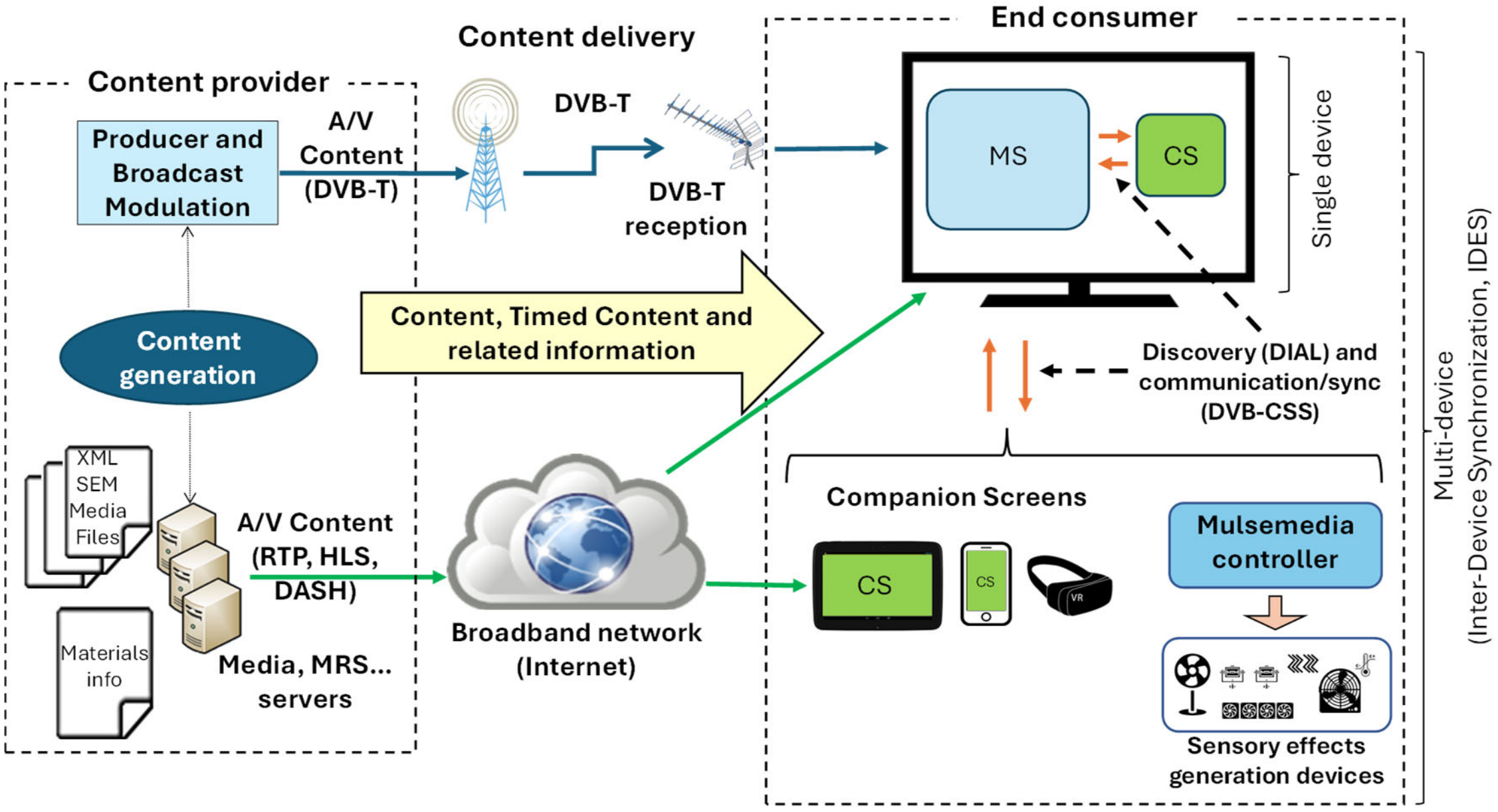

*Figure 11. Overview of the proposed end-to-end solution to include multi-sensorial effects in synchronized immersive hybrid TV scenarios.*

The Content Provider side (left-hand side of Figure 11) is responsible for, on the one hand, the broadcast AV media content delivered through DVB technologies (or through the Internet), and, on the other hand, for the generation, signaling, and storage of extra content, including the mulsemedia content (i.e., SEM metadata files) in media content servers available through the Internet. It is assumed that any metadata (e.g., specific MPEG2-TS tables) and content timelines (e.g., for sync purposes) are generated by the content provider at the time of content creation. Also, in mulsemedia TV services, which this proposal is focused on, for each video clip to be broadcast (main broadcast TV content), a Sensory Effect Metadata (SEM) file is created (to associate specific sensory effects with different temporal instances of that content). It is possible that content providers may support multiple delivery methods: terrestrial and satellite broadcast (i.e., DVB-T or DVB-S, respectively) or IP-based catch-up services using RTP, MPEG-DASH or HLS.

In the End Consumer part (right-hand side of Figure 11), the HbbTV-compliant TV Device[6] plays the main broadcast TV content (or even both broadcast and complementary broadband content), while, optionally, one or more companion devices[7], such as tablets or smartphones, run one or more companion screen applications (CS Apps) that will be part of the enhanced media content consumption experience. In the presented approach, the consumption of mulsemedia content also requires the management and synchronous activation of the available sensory effects at home, as specified in the SEM file. For this purpose, an additional device, named *Mulsemedia Controller (MC)* device, is employed to receive mulsemedia-related information from the main TV Device and to transmit the corresponding commands to the effect generation devices, thereby triggering the appropriate sensory effects in a synchronized manner. This device runs the named *Mulsemedia Controller Application* (*MC App*).

In companion screen experiences, content flows from the provider to the consumer, potentially in both timed and untimed formats. This flow of information may be augmented by additional information (or metadata) within the same TV (i.e., a hybrid terminal) or by *companion services* that may also be offered by the content provider (in the presented solution, the *Material Resolution Service*, MRS, defined in [12], is used as a Companion Service [11]). This additional information may be carried by different delivery technologies (e.g., where companion services are provided over the Internet and the timed content is delivered via terrestrial broadcast). Finally, the consumer uses these flows: the TV Device (running the MS App) consumes the timed content and the additional information (including sensory effects information needed for the mulsemedia experience), while the CS Apps consume other content flows (which may also be timed) needed for the companion experience (when available).

## 3.2 Content Provider Side

The main processes and components of the content provider side are summarized in this section.

### 3.2.1 Content generation

To provide multisensory hybrid TV experiences, the main AV broadcast content needs to be prepared to include or provide timelines and metadata. Moreover, the associated SEM file should have been created in advance and be accessible through broadband technologies.

#### *3.2.1.1 Broadcast content preparation*

A *Timeline* is the reference frame for measuring the progress of time for a given media stream [1]. In the proposed solution, among the types of timelines defined in [1], TEMI [27] has been chosen (following the same approach as in the authors' previous work in [5] , [34] or [35]). So, the DVB broadcast content (MPEG-2 TS) should include a TEMI timeline. Moreover, the URL of the MS App should be included in the MPEG-2 TS AIT (*Application Information Table*). Additionally, to use the IDES functions, a '*Stream Identifier Descriptor'* associated with the video stream must be added to the MPEG-2 TS PMT (*Program Map* Table). This is necessary for the TV Device to find the TEMI timeline associated with the broadcast video content. Furthermore, an event should be included in the MPEG-2 TS EIT (*Event Information Table*) to facilitate the generation of the *Content Identifier* (CI[8]) for the broadcast content.

---

[6] The TV Device is either a television or a device such as a set-top box (STB) that is connected to a TV. It is assumed that the TV device is connected to the home network and receives content using one of the above-mentioned delivery formats.

[7] The companion device is a generic term for a device owned by the consumer, such as (but not limited to) a smartphone or a tablet, which is used by the consumer whilst consuming complementary content. It is typically connected to the home network and, thence, to the Internet.

In the content prepared for the assessment, an MP4 video clip (encoded with H.264 for video and AAC for audio) was converted to an MPEG-2 TS that included that information using the *GPAC*[9] and *TSDuck*[10] tools. The former was used to insert the TEMI timeline, while the latter was used to modify the three mentioned tables.

##### *3.2.1.2 SEM metadata file generation*

The associated mulsemedia content for the main broadcast TV content is an SEM file, which should follow the SEDL schema (defined in [16] and included in the MPEG-V standard [9]). It uses tags to link a specific sensory effect and its parameters (e.g., type of effect, intensity, or other sensory effect-specific attributes) to a specific instant in the main broadcast TV content. In the proposed solution, both the main TV content and the additional mulsemedia content involved in the TV consumption experience will share a common timeline, enabling accurate sync at the consumer's home. To generate the XML-based SEM metadata files, any MPEG-V-compliant tool mentioned in section 2.2 can be used.

### 3.2.2 Companion Services

The solution proposed uses two external services or '*Companion Services*' [11] (Figure 13), the *Media Content and APP Service* and the *Material Resolution Service (MRS)* [12], which are briefly described in this section.

#### *3.2.2.1 Media Content and App Server*

A *Media Content and APP server* (e.g., a simple web or HTTP server) will, via broadband technologies (e.g., the Internet), provide additional media content, when available, complementary to the main broadcast TV content. Some examples include videos with additional related information, other camera views, related 360 videos, subtitles in different languages, audio guides, and the SEM metadata file associated with the main broadcast TV content. On the other hand, this server will also allocate the HTML5-based HbbTV MS App.

#### *3.2.2.2 MRS Server*

An important requirement in multisensory hybrid TV systems is the signaling of the presence of SEM metadata associated with the main broadcast TV content. In the proposed solution, a simple *MRS server* is needed to resolve material information based on provided content identifiers. This server listens for HTTP requests on a specific port and expects each request to include a parameter that contains the identifier of the broadcast content being watched on the TV Device (*Content Identifier*, or CI). The server will search for and provide an appropriate set of information for that CI, then return it to the client. This information is known as *Material Information*, and describes the following:

- the editorial structure of the broadcast, including proprietary identifiers and private data (including associated SEM metadata file location -i.e., its URL-).
- the timelines[11] signaled in the broadcast.
- how timelines signaled in the broadcast map to the timelines of the editorial structure.
- details of *Trigger Events* that may be signaled in the broadcast.
- URLs for endpoints that can provide updates to any of this information.

---

[8] The CI uniquely identifies the source of the timed content (e.g., the broadcast service the TV Device is tuned to) and other metadata coarsely identifying the editorial content being presented. For a broadcast, this coarse metadata includes the most recently signaled TV Program (DVB event) that is being presented for that service.

[9] https://gpac.io/ (last access: November 2026)

[10] https://tsduck.io/ (last access: November 2026)

According to [11], the broadcaster should implement the *CSS-MRS interface* with the following functions:

- Retrieving Material Information from the MRS.
- Providing updates to the Material Information.
- Providing correlation timestamps (the mapping between defined timelines).
- Providing updates to correlation timestamps.
- Providing private data (when needed).

The CI is a composite identifier that uniquely identifies content to the resolution of an individual program or presentation. It is derived from information that identifies the delivery stream and information contained within it. For the multisensory hybrid TV use case (a DVB service carried in a broadcast or IPTV-streamed MPEG2 TS), the format of the CI is described in clause 5.2.3 of [36].

The metadata for the materials associated with each content item are stored on the MRS server in a file that maps CIs to their corresponding material information. This file is called the *MRS file*. The format for material information included in the file is described in sections 7.4 and A.1.3 of [12]. As an example, Figure 12 shows the MRS file for the mulsemedia material that was used in the assessment of the proposal. It includes the URL of the SEM file (‘*demo.xml*’) associated with the broadcast content with CI ‘*dvb://0000.0000.0001;0002~20251215T1000Z--PT00H17M*’, and timing information (timelines and mapping).

## 3.3 Consumer Side

Figure 13 and Figure 14 provide an overview of the consumer-side architecture of the proposed solution, focusing on the internal architecture of the HbbTV-compliant TV Device and the MS App running on it, and of the MC App running on the MC device, respectively. From an architectural perspective, the TV Device, the MS App, and the MC App each contain several functional blocks, termed elementary functions, that handle interface interactions, as explained in the subsections below.

### 3.3.1 HbbTV-compliant TV Device – MS App

The TV Device should be compliant with the HbbTV standard [1] and with the ETSI TS 103 286-2 specification [12]. As shown in Figure 13, the TV Device includes some elementary functions (CII, App2App, Wall Clock or WC, and Media Sync Application or MSAS services) and interfaces (CSS-CII, App2App, CSS-WC and CSS-TS) defined in those specifications. It also runs the HbbTV MS App, which is associated with the main broadcast TV content. During the multisensory hybrid TV session, the MS App transmits sensory-effect activation commands to the MC App, based on the main broadcast TV content playout progress on the TV Device.

[11] Timelines are described in terms of their tick rate, and the mappings between timelines are described by pairs of values (one from each timeline) known as *Correlation Timestamps*.

The TV Device receives broadcast content from a broadcaster, including metadata that coarsely identifies the editorial content. Through the CSS-CII interface, it should provide the MS App with the Content Identifier (CI) of the broadcast content being watched on TV. It should also enable the MC App to use a timeline to synchronize the generation of sensory effects with the main broadcast TV content playout on the TV Device. For this purpose, it should provide the MC App (again, through the CSS-CII interface) with the location of an endpoint for implementing the CSS-WC and CSS-TS interfaces. As shown in Figure 13, the CSS-WC interface is used by the WC Server of the TV Device and the WC Client in the MC App. In contrast, the CSS-TS interface is used by the MSAS Server of the TV Device and the Sync Client (SC) in the MC App for their elementary functions.

```
"dvb://0000.0000.0001;0002~20251215T1000Z--PT00H17M": {
  "type": "response",
  "version": "1.1",
  "rev": "1",
  "repollingInterval": 0,
  "materials": {
    "mulsemedia_file": {
      "ids": [
        {
          "type": "urn:url",
          "id": "https://IP_addr_WebServer/material/mulsemedia/demo.xml"
        }
      ],
    }
  },
  "syncTimelineInformation": [
    {
      "contentIdStem": "dvb://0000.0000.0001",
      "timelineSelector": "urn:dvb:css:timeline:temi:8:1",
      "timelineProperties": {
        "unitsPerTick": 1,
        "unitsPerSecond": 12288,
        "accuracy": 0.0
      },
      "mappings": [
        {
          "materialIndex": "mulsemedia_file",
          "start": "0",
          "end": "9000000",
          "correlationsChanging": false,
          "correlations": [
            {
              "materialPoint": "0",
              "point": "0"
            }
          ]
        }
      ]
    }
  ]
}
```

*Figure 12. Example of a JSON file including material information*

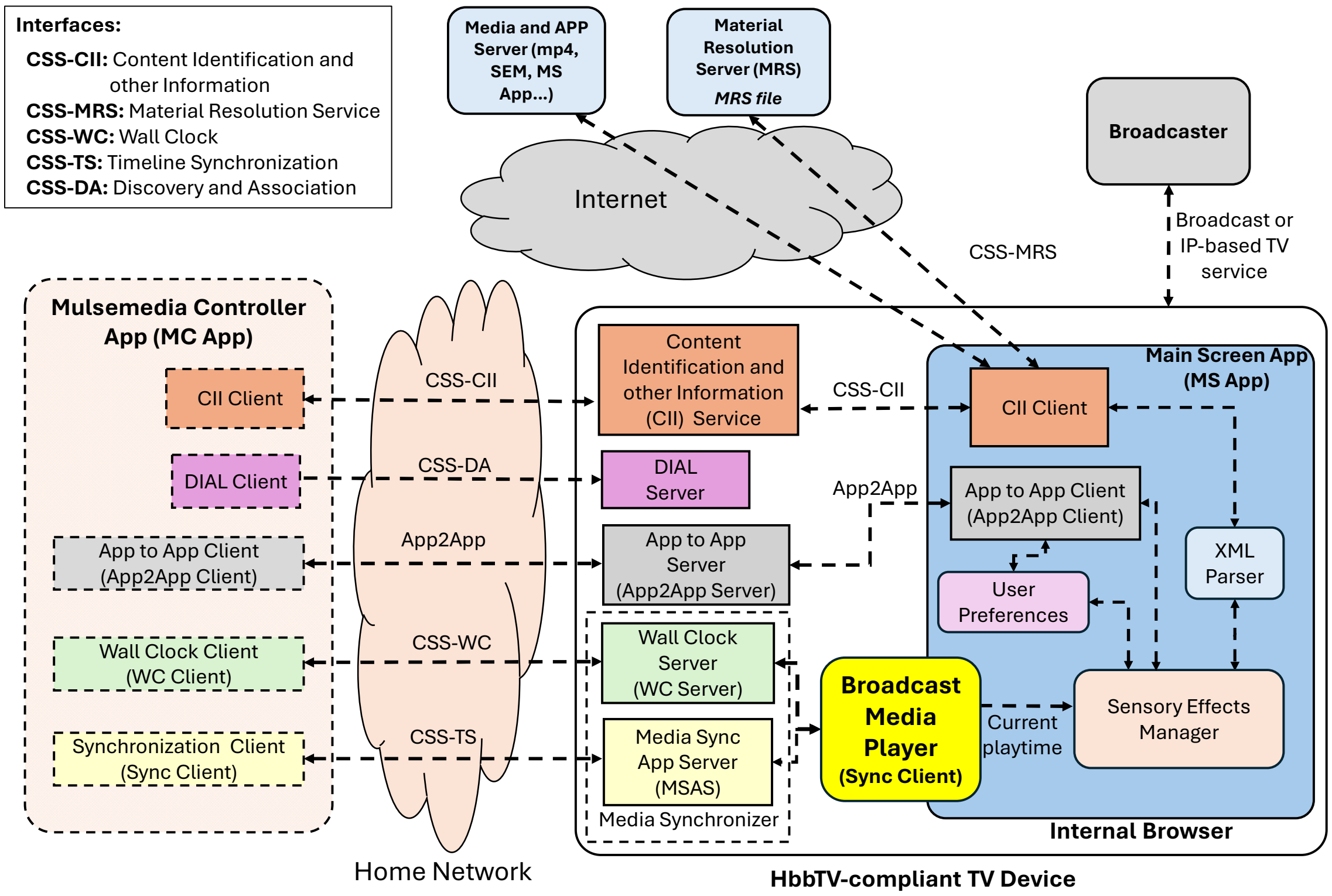

*Figure 13. General architecture – TV Device side*

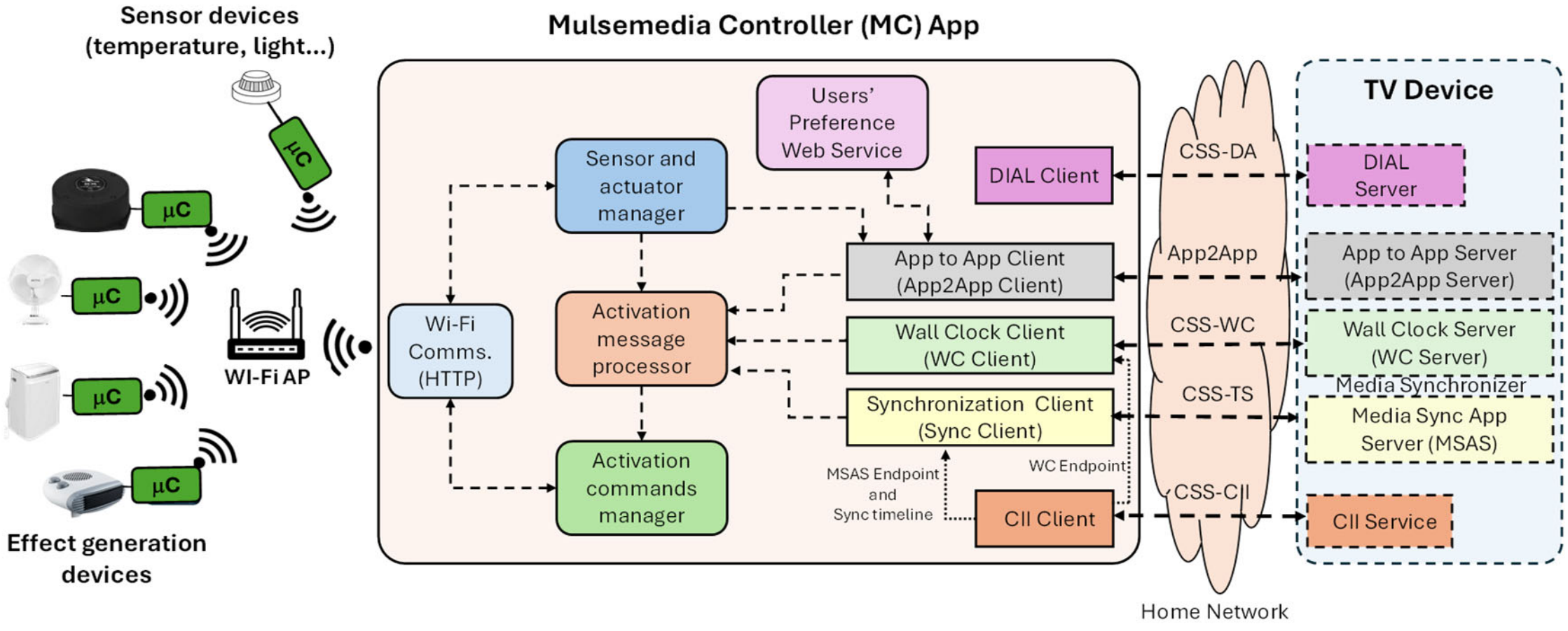

*Figure 14. General architecture – MC Device side*

The App2App Server on the TV Device enables a connection (i.e., a channel) between the App2App Clients of both the MC and MS Apps to exchange MPEG-V-compliant messages. This server provides the same functionality as the one provided by the WebSocket Server in the authors' previous mulsemedia platform presented in [10].

### 3.3.1.1 MS APP in the TV Device

This section provides a brief overview of the MS App running on the TV Device. It should be implemented using HTML5 technology (HTML, CSS, and JavaScript). The HbbTV *video/broadcast* embedded object enables interaction with and control of the TV Device's underlying broadcast and video playback capabilities. This object provides the MS App with a standardized interface for accessing the linear main broadcast TV content, managing media playback, and retrieving metadata.

#### 3.3.1.1.1 MS App Architecture

Figure 13 illustrates the architecture of the MS App (blue square) running on the TV Device, which is composed of several functions or modules explained below:

- *CII Client*. At the start, the MS App needs to know the broadcast Content Identifier (CI) and the TV Device's presentation status. The CII client can obtain the location of the service endpoint (i.e., the URL for the TV *Device Content Identification and other Information -CII- Service*) for the CSS-CII interface, which is implemented internally by the HbbTV-compliant TV Device. The CII client retrieves, via that interface, the CI for the timed broadcast content being presented by the TV Device and the URL of the CSS-MRS service endpoint (i.e., the URL of the MRS Server). Then, it sends a request with the CI to the CSS-MRS service endpoint[12] via the CSS-MRS interface and retrieves the broadcast content metadata (i.e., *Materials*) in JSON format. As shown in Figure 12, the returned information includes a download link for the associated SEM metadata file, along with timing information (timelines, mappings, etc.) for the broadcast content being watched on TV. The SEM file is then fetched and passed to the *XML Parser* module.
- *XML Parser*. This module receives XML-based information (SEM metadata, the generation device's capabilities, or the user's preferences) to be processed. The data obtained are stored in appropriate internal data structures (i.e., variables) for use by the application.
- *Sensory Effects Manager*. The internal data structures generated by the previous module are passed to this module, which also receives information about the playout point of the TV Device main broadcast TV content player. This module is responsible for checking whether, at any playout instant, sensory effects need to be activated/deactivated at a specific location in the local scenario. If so, it will send this information to the App2App communications client, allowing the corresponding activation/deactivation message to be sent to the MC App via a previously established channel through the internal App2App Server connecting both apps. To do this, it creates MPEG-V-compliant activation/deactivation messages with appropriate parameters (based on the generation device's capabilities and the user's preferences) and passes them to the App2App communications client module.
- *App2App communication client*. This module establishes a connection (i.e., a channel) with the MC App, via the TV Device's App2App Server, to exchange messages containing MPEG-V standard-compliant XML data.
- *User's preferences Module*. In personalized mulsemedia TV experiences, the users can choose which sensory effects, from among those available at their homes, they want to be activated or not during the session, discarding those they do not want. Additionally, they can configure the properties of all available sensory effects (e.g., the maximum or minimum intensity of the effect when generated during the experience). If users configure their preferences, only messages corresponding to the activation/deactivation of the desired sensory effects, as defined by their configured properties, will be passed to the App2App communications client module for transmission to the MC App.

---

[12] In the proposal, it is expected that the MS App is provided by the broadcaster and, therefore, knows the endpoint's location. Alternatively, the endpoint of the CSS-MRS interface could be signaled in the broadcast presented by the TV Device and then passed to the MS App via the CSS-CII interface (how to do this is outside the scope of this paper).

#### 3.3.1.1.2 Management of the activation/deactivation of sensory effects.

During the multisensory hybrid TV session, the MS App considers the sensory effects defined in the SEM metadata, their timing information, the capabilities of the available sensory effects generation devices in the local scenario, and the configured user's preferences. Depending on the main broadcast TV content playout instant provided by the TV Device, the devices' capabilities, and the user's preferences, it continuously decides when to send MPEG-V standard-compliant activation/deactivation messages to the MC App via the established App2App communications channel. On the one hand, the MC App requires time to process these messages, translate them into the corresponding activation/deactivation commands, and send them to microcontrollers managing the sensory-effect generation devices, ensuring they reach those devices on time. On the other hand, the sensory effects generation devices must receive commands in advance to have time to start up (or warm up) and to provide the sensory effect at the appropriate intensity at the right time (i.e., synchronized with the playout of the main broadcast TV content). The collected capabilities of all the sensory effects generation devices include the *zeroth order delay time*[13] and the *first order delay time*[14] properties. The MS App should perform the necessary calculations (considering the capabilities of each generation device and the mean measured end-to-end delays) to send activation/deactivation messages to the MC App, along with timestamps, with sufficient advance notice. More details about this can be found in [10].

### 3.3.2 MC Device – MC App

The primary function of the proposed MC App running on the MC device is to provide a communication gateway between the MS App and the microcontroller devices that manage the sensory effects generation devices available in the local mulsemedia scenario (see [10]). Notice that generation devices for all the sensory effects required by the MS App (i.e., for all the effects defined in the SEM metadata of the selected AV media content) may or may not be available in the local multisensory TV scenario. The MC App is responsible for discovering all sensory effect generation devices in the local environment, detecting any changes in their status (e.g., they may be switched off or lose network connectivity), and notifying the MS App. Additionally, the MC App must process and interpret MPEG-V standard-compliant activation/deactivation messages received from the MS App, and translate them into corresponding command messages to be sent to the microcontroller devices. The MC App architecture and its basic operation are described below. Then, the various processes involved in or interacting with it for the correct performance of the proposed solution are explained in detail.

#### *3.3.2.1 Application architecture*

The MC App also features a modular architecture, as illustrated in Figure 14, with the following modules:

- *DIAL Client*. This module detects the presence of an HbbTV-compliant TV Device on the home network.

[13] *Zeroth order delay* time is the required preparation time of a sensory device to be activated since it receives a command.

[14] *First order delay* time is the required time for a device to reach the target intensity since it receives a command and is activated.

- *CII Client*. It is used to retrieve the WC and MSAS service endpoints from the TV Device's CII Service.
- *App2App Communication Client.* It is responsible for communication with the MS App (e.g., via the App2App Server on the TV Device).
- *Users' preferences server module.* It provides a web service to configure and store the user's profile and preferences in the system.
- *Sensors and generation devices manager.* This module is responsible for detecting the presence of sensors and sensory effects generation devices connected to the Wi-Fi private network in the local mulsemedia scenario. It obtains their capabilities and passes them to the MS App via the App2App communication client. For sensors, this module collects and transmits their measurements to the MS App.
- *Sensory effect activation message processor.* It is responsible for processing MPEG-V standard-compliant XML-based activation/deactivation messages received from the MS App, and passing the information to the next module.
- *Activation commands manager.* It receives information from the previous one about which real sensory effects to activate or deactivate and translates it into the corresponding activation/deactivation commands to be sent via the *Wi-Fi Communications module* to the corresponding microcontroller devices.
- *Wi-Fi Communications module.* It communicates with microcontroller devices over the local Wi-Fi network.
- *WC and Sync client.* Both are used for IDES purposes, as explained below.

#### *3.3.2.2 Discovery and services endpoints retrieval*

Using the DIAL protocol [6], the MC App can discover the TV Device and then retrieve the App2App and CSS_CII service endpoints (i.e., their URLs), as explained in section 14.7 of [1]. Then, it also retrieves the WC service and MSAS service endpoints from the CII Service.

#### *3.3.2.3 User's preferences service*

To allow for personalized experiences, users can configure their preferences for which sensory effects to include (or not) in their multisensory hybrid TV sessions, as well as the characteristics of those effects. The MC App includes a web service module that allows adjusting the configuration of the sensory effect generation devices, taking these preferences into account. For example, a pregnant woman may be intolerant of certain smells or prefer to avoid annoying vibrations or sudden movements, so she can exclude any scents that might bother her and adjust the vibration settings (or its main intensity) accordingly. The user's preferences configuration web page (Figure 15) consists mainly of a scrollable window with configurable settings for each available sensory effect generation device in the local mulsemedia scenario, based on their characteristics and positions within it. User's preferences are saved on the MC device in an XML file, so that the MS App can retrieve them in future sessions. The MPEG-V standard only allows configuring the user's preferences for each type of sensory effect. However, the proposed service enables users to configure their preferences for each sensory effect generation device in the local mulsemedia scenario. This may be necessary in situations where a particular sensory effect from a specific position can be annoying for some users (e.g., when there are several fans and the user wants to reduce the maximum airflow intensity of only the closest one). The MPEG-V specifications were taken as a starting point, but the base attributes of the sensory effect preferences (*userSensoryPreferenceBaseAttributes*) were extended by adding a new attribute of type *anyURI*, called *deviceIdRef*, which is included in all of them. This allows associating each user's preference for a specific sensory effect (*USPreference*) with each generation device that produces that effect in the scenario. Figure 16 shows an example of a file containing the user's preferences shown in Figure 15, for a wind generation device (location *right:bottom:front*), which is disabled (*activate='false'*), and another for a scent generation device (location *centerright:middle:front*), enabled but with 3 blocked scents (*banana, pastry and coffee*) and 4 unblocked ones: *forest, strawberry, orange and chocolate*.

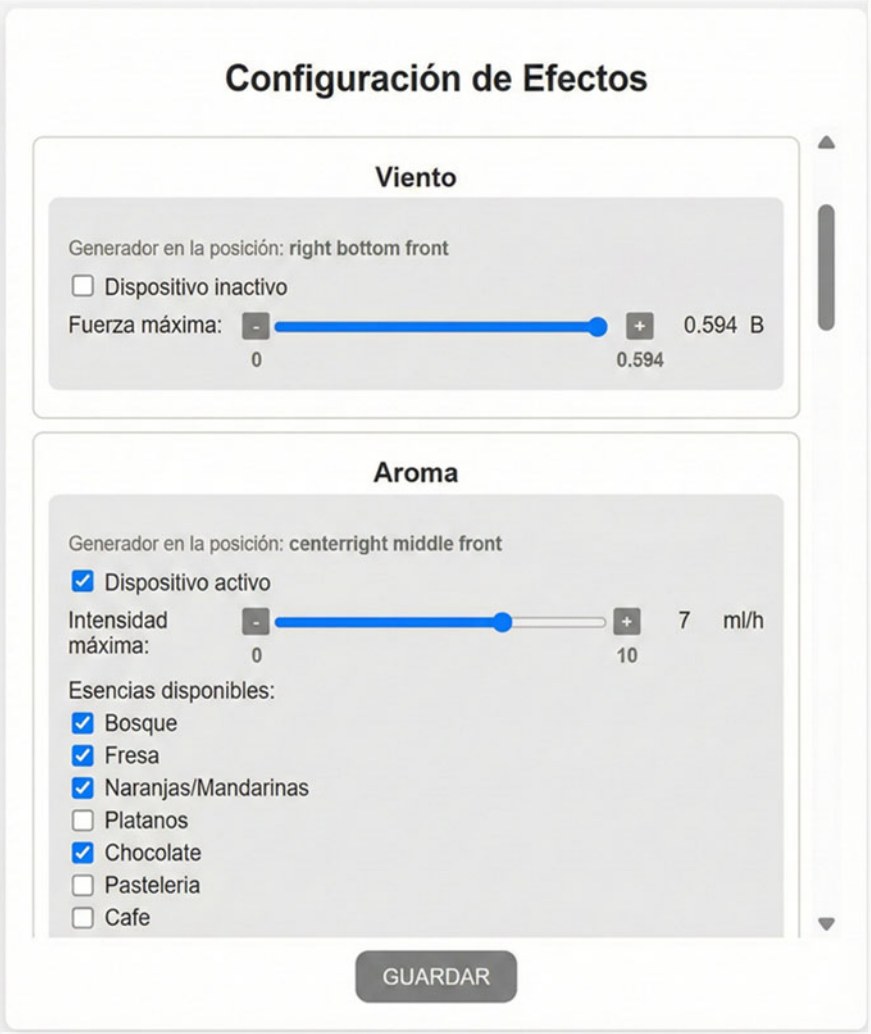

*Figure 15. User's preferences configuration web page. Example of configuration of wind and scent effects generation devices in two specific locations*

### 3.3.2.4 Application operation

When the MC App is launched, its DIAL client first finds the TV device (as described above). Then, the App2App communications client establishes a connection to the TV's App2App communications server, which serves as the communication channel between the App2App clients of both MC and MS Apps. It also initiates a process, described in [10], to discover sensor or effect generation devices within the Wi-Fi local network. Through this discovery process, the MC identifies the available sensory effects and their primary characteristics, storing them in a list. In this list, for each sensory effect present, the sensory effect type, the position from which it is activated, and the capabilities of its generation device are stored. This discovery process will be repeated periodically to identify new devices or detect old ones that may have been lost or removed.

When the MS App first connects to the MC App, after establishing the App2App communications channel, it will request this information, which will be sent using an XML-based MPEG-V standard-compliant message (see [10]). In addition, if there are already stored users' preferences, it will also send another XML-based MPEG-V standard-compliant message providing them. From that moment on, the MC App will only send messages to the MS App when changes in the local environment and/or the stored user's preferences are detected or when it receives data from the sensors present in the local mulsemedia scenario.

When the MC App receives an activation/deactivation timestamped message for a sensory effect from the MS App, it processes and translates it into the corresponding command message, which it will send over the Wi-Fi network at the appropriate instant to the microcontroller device responsible for generating that sensory effect in sync with the AV media content playout.

All the processes, communication protocols and message exchange mentioned in this section are explained in [10].

```
<?xml version="1.0" encoding="UTF-8"?>

    <cidl:ControlInfo
        xmlns:xsi="http://www.w3.org/2001/XMLSchema-instance"
        xsi:schemaLocation="urn:mpeg:mpeg-v:2012:01-CIDL-NS CIDL.xsd"
        xmlns:cidl="urn: mpeg:mpeg-v:2012:01-CIDL-NS"
        xmlns:dcdv="urn: mpeg: mpeg-v:2012:01-DCDV-NS"
        xmlns:sapv="urn: mpeg: mpeg-v: 2012:01-SAPV-NS"
        xmlns:scdv="urn:mpeg:mpeg-v:2012:01-SCDV-NS"
        xmlns:sepv="urn:mpeg:mpeg-v:2012:01-SEPV-NS"
        xmlns:mpegvct="urn: mpeg:mpeg-v:2012:01-CT-NS">

        <cidl:UserSensoryPreferenceList>

            <cidl:USPreference xsi:type="sepv: WindPrefType"
                deviceIdRef="wind1"
                activate="false">
            </cidl:USPreference>

            <cidl:USPreference xsi:type="sepv: ScentPrefType"
                deviceIdRef="Scent1"
                maxIntensity="7"
                activate="true">

                <sepv: HedonicScent>
                    :mpeg:mpeg-v: 01-SI-ScentCS-NS: Platanos
                </sepv: HedonicScent>

                <sepv: HedonicScent>
                    :mpeg:mpeg-v: 01-SI-ScentCS-NS: Pasteleria
                </sepv: HedonicScent>

                <sepv: HedonicScent>
                    :mpeg:mpeg-v: 01-SI-ScentCS-NS: Cafe
                </sepv: HedonicScent>
            </cidl:USPreference>

        </cidl:UserSensoryPreferenceList>

    </cidl:ControlInfo>return go(f, seed, [])
}
```

*Figure 16. Example of user's preferences regarding two generation devices with deviceIdRef 'wind1' and 'scent1'*

### 3.3.3 Inter-Device Synchronization (IDES)

The HbbTV-compliant TV Device implements interfaces and protocols defined in [12]. The IDES functionality is controlled by the HbbTV MS App through the *MediaSynchroniser* object defined in clause 8.2.3 of [1]. This object performs the IDES-related elementary functions defined in [12]. The *MediaSynchroniser* serves as the Sync Client (SC) to control the timing of the presentation of the broadcast video content (also known as the *sync media*) and to communicate timing information to the MSAS function. Those tasks are implemented by the *MediaSynchroniser* function in the HbbTV-compliant TV device. The *MediaSynchroniser's* functions in the TV device communicate with the MC App using the CSS-CII, CSS-WC and CSS-TS protocols defined in [12], as shown in Figure 13 and Figure 14. The TV device therefore implements WC server, MSAS and SC functions. In it, the service endpoints for these protocols are provided by the *MediaSynchronizer's* functions, i.e., a DVB CSS-CII endpoint, a DVB CSS-WC endpoint, and a DVB CSS-TS endpoint.

The reference point for the generation and interpretation of timestamps by the HbbTV-compliant TV device and the MC App used in IDES should be implemented as defined in clause 5.7.2 of [12].

For IDES purposes, processes for synchronizing the clocks and timelines of the MS and MC Apps are needed and are explained below.

#### *3.3.3.1 Wall Clock Service. Common time reference*

The HbbTV-compliant TV Device shall have a local time source, known as the *Wall Clock* (WC), as defined in clause 8.3 of [12], which shall be monotonic and without discontinuities. Moreover, the device shall implement a WC-Server as defined in clause 8 of [12]. The WC-Server shall provide the CSS-WC service endpoint on the terminal's broadband interface.

In the proposal, to provide a common time reference for the MC App, the WC service elementary function included in the TV Device is used. The WC Server has access to the Wall Clock, and the measurements of it by the TV Device shall have a measurement precision of 1 ms or better (smaller) [12]. The WC Client elementary function of the MC App communicates with the WC server on the TV Device through the CSS-WC interface (see Figure 14) to estimate the WC on the TV Device. Then, the WC Client of the MC App will synchronize its WC estimate with the WC of the WC Server (i.e., the TV Device). For this purpose, messages are exchanged between the WC Server and the WC Client, and time values are sampled from both sources. Using them, the WC Client can estimate the TV Device WC progress while compensating for network latency in the exchange of protocol messages.

#### *3.3.3.2 Timeline synchronization*

The MC App needs to generate the sensory effects in sync with the timed content being presented on the TV Device. To facilitate IDES, the HbbTV-compliant TV Devices implement the CSS-TS service endpoint (as defined in clause 9 of [12]). The MC App connects to that endpoint via the CSS-TS interface to establish a *Timeline Sync Protocol* session. This protocol conveys messages containing *setup data* and *Control Timestamps and Actual*, *Earliest* and *Latest Presentation Timestamps* that relate WC time to the timeline of the broadcast content watched on the TV Device (termed the *Sync Timeline*).

A timeline sync process is needed through the CSS-TS interface. The MC App needs a means to synchronize itself against the timeline of the broadcast TV content being watched on the TV Device. For this purpose, the HbbTV-compliant TV Device shall, on the one hand, implement the *Media Sync Application Server* (MSAS) elementary function, which provides a CSS-TS service endpoint. For the broadcast media object involved in sync, the TV Device implements a *Sync Client* (*SC*) function that interacts with the MSAS function to control the presentation timing and report the achievable presentation timings for that media object. On the other hand, the MC App also includes an SC module or function implemented in the broadcast media player (see Figure 14, in yellow).

The MSAS function of the TV Device sends control timestamps to the SC of the MC App, to be considered when generating sensory effects synchronized with the playout of the broadcast TV content watched on the TV Device. Those control timestamps shall represent the timing of presentation of the called *sync media* (i.e., the broadcast content) by the TV Device.

The Timeline Sync between the SC of the MC App and the MSAS is done through the CSS-TS interface, comprising:

- the SC function using the *Content Identification and other Information (CII) Service* to obtain the *Sync Timeline***.** It is used as the reference frame for exchanging timing information across the CSS-TS interface using the CSS-TS protocol (as described in clauses 5.7 and 9 of [12]). The Sync Timeline defines the reference frame for *contentTime* property values in *Control Timestamps and Actual, Earliest* and *Latest Presentation Timestamps* exchanged during the protocol session between the MS App on the TV Device and the MC App on the MC device.
- the MSAS function and SC module exchanging presentation timing information in terms of the selected timeline.

This way, the MC App can synchronize the generation of sensory effects with the timed content being watched on the TV Device.

Timeline Sync is an ongoing process that involves regular communication, whilst sync is to be maintained to avoid drift and cope with changes in presentation timing by the TV Device (such as discontinuities due to the user using a pause function on the TV Device).

When timing information (for timeline sync) is exchanged between the TV Device (i.e., its MS App) and the MC App via the CSS-TS interface, the messages conveying it will be subject to network latency. For the MC App to achieve accurate timeline sync with the TV Device, both the MC App and the TV Device must compensate for the latency. The exchanged timing information for timeline sync is expressed in terms of the WC. The MC App, upon receiving a message, can observe the WC's apparent progress since the message was generated and therefore compensate for network latency. For example, the TV Device sends a message across the CSS-TS interface to the MC App indicating that the timeline of the timed content it is presenting was at time X when the WC was at time Y. For simplicity, in this example, the units (the tick rate) of both the WC and the Timeline are assumed to be the same. The MC App receiving the message observes that its own WC is now at time Y+D, where D is the difference between the WC time Y, when the Timeline was observed to be at time X, and the WC time at which the MC App makes this observation. The MC App, therefore, knows that the current point on the timeline for the timed content presented by the TV Device is X+D. Notice that the accuracy with which the WC is estimated by the WC Client limits the achievable sync accuracy between the TV Device and the MC App. The accuracy of the estimate should be maximized by using stable, accurate clock oscillators, maximizing clock measurement precision, and minimizing the latency of exchanging CSS-WC messages.

# 4 Mulsemedia TV prototype

As proof of concept for the proposed approach and to enable subjective validation by users, a prototype of a real-life multisensory TV scenario has been implemented in a research laboratory at the Gandia Campus of the Polytechnic University of Valencia (Spain). The end-to-end platform for hybrid content generation, delivery, and consumption used in some of the authors' previous works ([5], [34] or [35]) has been extended, enhanced, and combined with the mulsemedia platform presented in [10] to create an end-to-end multisensory hybrid TV testbed. Some new services (e.g., MRS) have been included, and a few software modifications have been made to comply with the HbbTV specification. The prototype overview is shown in Figure 11.

## 4.1 Hardware

For content broadcast, in the content provider part, a desktop Personal Computer (Intel Core i7-6700 @ 3.40GHz CPU, 8GB RAM, Windows 10) equipped with a DekTek DTA-2111 0-1GHz PCI Express modulator card has been used.

For running the MC App and the required external servers (Media Content and App Server, and MRS Server), a single Intel NUC11TNHi5 mini-computer (MiniPC) was used, featuring 16.0 GiB of RAM, an 11th Gen Intel Core i5-1135G7 processor, Intel Xe Graphics (TGL GT2) GPU, and a 1 TB hard drive. It acts as the MC device and, for the prototype's simplicity, also hosts both external servers.

In the consumer part, for the TV content consumption with multisensory effects, the authors' web-based MPEG-V standard-compliant platform presented in [10] has been used. In this part, a typical small living room has been recreated, where a user can enjoy an immersive mulsemedia consumption experience. It is a 3x3x3m space, with a TV on a typical piece of furniture and a 2-seater sofa in front of it, positioned so that the center of the TV is approximately 2.5 m from the users' heads while sitting on the sofa. Several off-the-shelf devices for generating sensory effects have been placed around the sofa at the positions and with the characteristics indicated in Table 1. Figure 17 illustrates the location of the generation devices in the scenario, while Figure 18 displays a photograph of the prototype's rear view. The details of the local mulsemedia platform architecture, the communication protocols for activating the sensory effects, and the involved generation devices can be found in [10]. In this case, since an HbbTV (v2.0.1 or higher)-compliant TV Device is needed for the proposed multisensory hybrid TV solution, the SONY device, model K50S35B.CEI (TV LED 50' - Sony Bravia 3, Direct LED, 4K) has been used.

## 4.2 Software

In this section, the employed software components and tools used in the prototype are described.

For the generation, from scratch, of the SEM metadata associated with the broadcast AV content used in the assessment described later, the SeVino2 tool ([18] [19]) has been used, along with some required manual modifications.

In a broadcast DVB-T environment, MPEG-TS is the standard container for digital television. So, the original video content, initially encoded in MP4 format (H.264 for video and AAC for audio), is transcoded to MPEG Transport Stream (MPEG-TS) using the *FFmpeg*[15] tool. Then, as explained in section 3.2.1.1, that stream needs to be prepared by using the *GPAC* and *TSDuck* tools. The former was used to insert the TEMI timeline, while the latter was used to modify its AIT, PMT, and EIT tables.

*Table 1.Off-the-shelf sensory effects generation devices included in the multisensory TV scenario*

| Sensory effect | Generation device (trademark and model) | Position (MPEG-V) |
|---|---|---|
| Wind | Fan Cecotec EnergySilence 1000 Pro | Left:top:front |
| Wind | Fan Cecotec EnergySilence 1000 Pro | Right:top:front |
| Wind | Fan Cecotec EnergySilence 1000 Pro | Left:top:back |
| Wind | Fan Cecotec EnergySilence 1000 Pro | Right:top:back |
| Vibration | Haptic Vest bHaptics TactSuit X40 | Center:middle:midway |
| Vibration | Two subwoofers Monacor BR-50 | Center:bottom:midway |
| Scent | Olorama, 10 scent diffuser | Centerleft:middle:front |
| Scent | Olorama, 10 scent diffuser | Centerright:middle:front |
| Lights | Smart Philips Hue Fugato + Hue Bridge | Center:top:back |
| Lights | Philips TV with Ambilight system | Center:middle:front |
| Flash | Mini LED Strobe Light LET'S ROCK | Center:top:back |
| Cooling | AA/CC Hisense APC12QC | Left:middle:front |
| Cooling | AA/CC Hisense APC12QC | Right:middle:front |
| Heating | Heater Taurus Tropicano 2400 | Left:middle:midway |
| Heating | Heater Taurus Tropicano 2400 | Right:middle:midway |

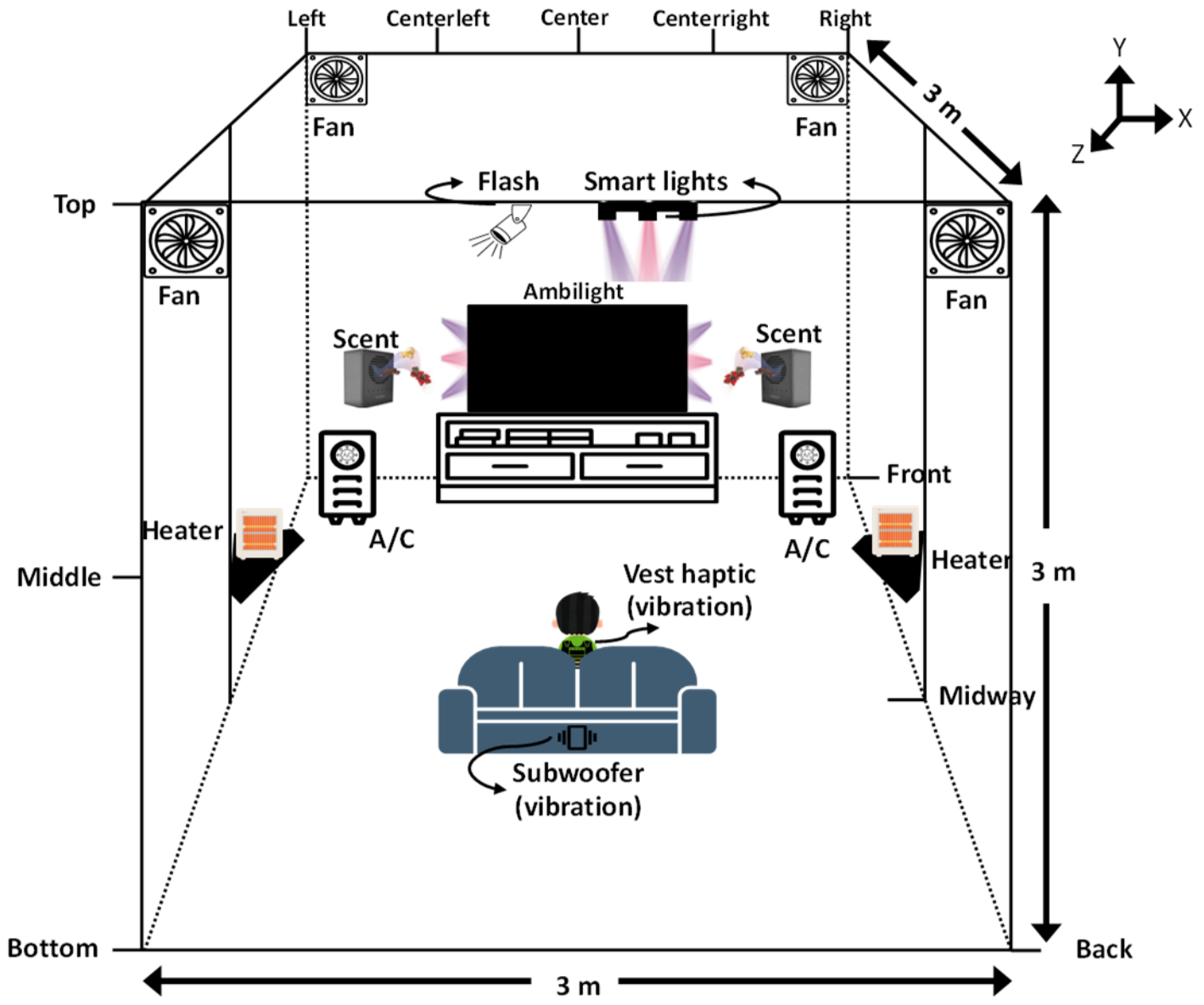

*Figure 17. Location of the sensory effects generation devices in the multisensory TV scenario*

---

[15] https://www.ffmpeg.org/ (last access: November 2026)

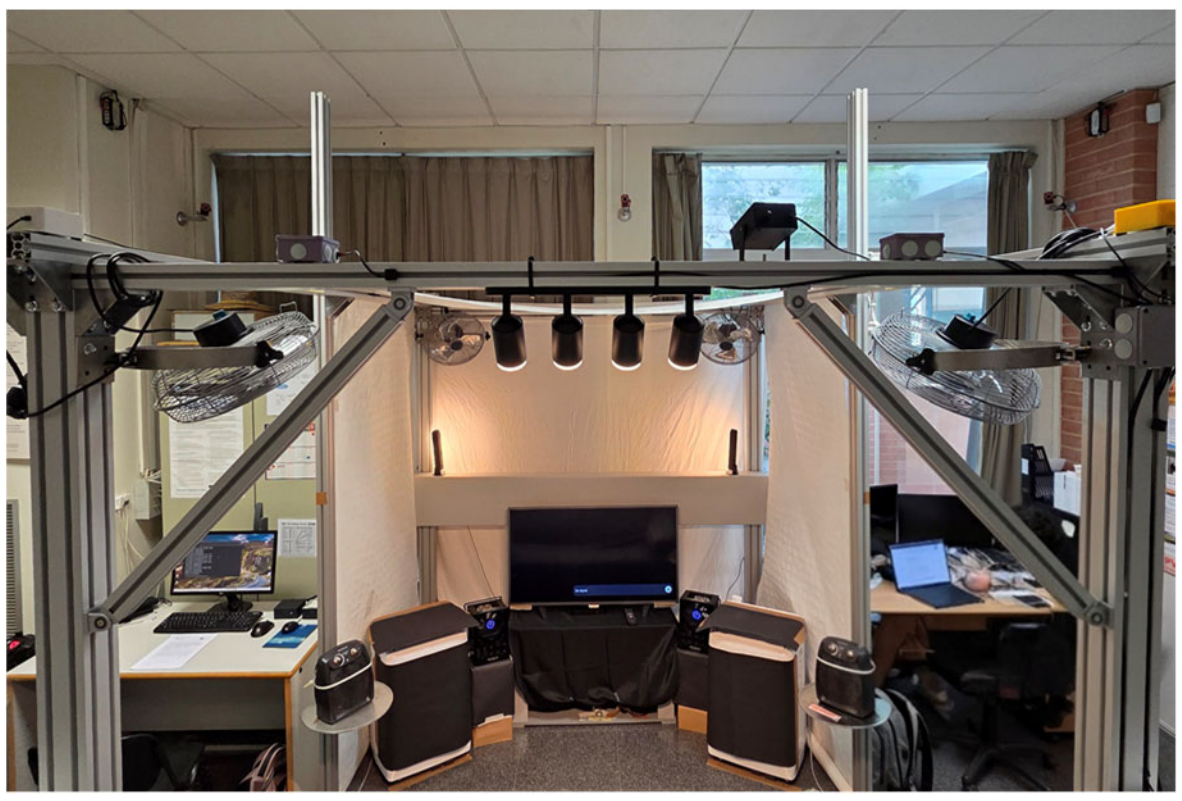

*Figure 18. Rear view of the implemented prototype of the multisensory TV scenario*

To broadcast the prepared MPEG2-TS, the Windows 10-based PC with the mentioned modulator card is used, employing the *DekTec StreamXpress* software provided by the card manufacturer. To accurately simulate a DVB-T-based DTT environment, a 64-QAM modulation scheme is configured.

A very simple MRS Server has been implemented from scratch in Rust to provide a JSON-based MRS file when required (see section 3.2.2.2). It is responsible for resolving material information based on provided content identifiers. It listens for requests on the specified port and expects each request to have a parameter named “*contentId*” The service then looks up the material information associated with that content ID and returns it in the response (see Figure 12).

On the consumer side, an HbbTV-compliant MS App has been developed following the architecture described in section 3.3.2.1.1 and the guidelines outlined in the HbbTV Developer portal, along with examples available on GitHub[16]. It is automatically launched when the TV Device reads its URL in the AIT Table of the received MPEG2-TS. The MS App is an HTML page without graphical elements (like a transparent web page) that runs only a few JavaScript scripts. Those scripts include most of the code from the original web-based MS App, which is included in the authors' mulsemedia platform described in [10], and has been reused in this new app. Unlike the MS App on the original mulsemedia platform, the newly implemented HbbTV-compliant one uses the HbbTV TV Device elementary functions and interfaces previously described, and makes use of the *MediaSynchroniser* object previously explained for timeline sync (i.e., for IDES purpose). The communication protocol, messages, and the interaction with the MC App to manage the activation/deactivation of sensory effects are the same as explained in [10], but, in this case, the App2App channel established through the TV Device App2App Server is used instead of a dedicated WebSocket Server, as in that work. The MS App is located on the (web) Media content and App Server running on the MC Device (miniPC). It includes a CII Client to obtain the required endpoints and the MRS file and provides it with the URL of the SEM file. On the other hand, the original MC App from [10] has been modified to adhere to the architecture proposed in section 3.3.2.1, in line with the main guidelines for CS apps in the HbbTV specification [1]. It includes a DIAL Client to discover the TV Device, a CII Client to obtain the required endpoints to communicate with the MS App, and WC and Sync Clients to achieve accurate IDES (per the DVB-CSS specification) for the broadcast content watched on the TV Device. App2App Clients of both apps can communicate through the MSAS service provided by the HbbTV-compliant TV Device to exchange messages related to sensory effects management (effect generation capabilities, user's preferences, activation/deactivation commands, etc.) during the mulsemedia hybrid TV session.

[16] https://github.com/HbbTV-Association/Tutorials (last access: November 2026)

# 5 Evaluation

This section presents the conducted subjective evaluation of a Mulsemedia TV experience. First, the AV media content used during the assessment sessions is described. Then, the participants, the methodology followed, the results obtained, and a brief discussion are presented. Finally, some of the study's identified limitations are summarized.

## 5.1 Selected contents

For the assessment, a 17:40-minute video clip[17] was created, the characteristics of which are summarised in Table 2. It was initially encoded in MP4 format (H.264 for video and AAC for audio), then transcoded to MPEG2 Transport Stream (MPEG2-TS) and prepared for the HbbTV-based service, using the software described above. Considering the work in [37], the video clip includes 6 different content genres selected for consumption during the assessment, such as news, sports, documentaries, commercials, cooking shows, and cinema. The video has 12 parts (Table 3), each with a different content genre, stitched together using transitions or short introductory animation sequences made with both the *CapCut* video editor[18] and the *Gemini VEO 3* AI-based generation tool[19]. The source of each part is indicated in the last column, and some URLs are also provided in footnotes. Regarding the source of the videos, two were obtained from the dataset in [18]; four were provided by À Punt Media[20], a Valencian regional TV; and the rest were collected from YouTube.

As already mentioned, the SeVino2 tool was used to generate a preliminary XML-based SEM metadata file from scratch, and a few additional manual modifications were made later. Only sensory effects available in the local mulsemedia platform employed in the assessment have been added. It includes all the effects available on that platform, most of which overlap or are very close, but without overloading the experience[21]. Appendix I provides the list of the sensory effects included in the SEM file associated with the video clip prepared for the assessment, along with their properties.

*Table 2. Features of the video clip selected for the assessment*

| Video | Resolution/FPS | Duration (mm:ss) | Video Bitrate (kbit/s) | Size (MB) | Audio Bitrate (kbits/s) | Audio Channels | Audio Sampling Rate (kHz) |
|---|---|---|---|---|---|---|---|
| 1 | 1920x1080/30 | 17:40 | 3738 | 494 | 159 | 2 (stereo) | 48 |

[17] https://youtu.be/ge9AF_FZIMc (last access: November 2026)
[18] https://www.capcut.com/ (last access: November 2026)
[19] https://aistudio.google.com/models/veo (last access: November 2026)
[20] https://www.apuntmedia.es/ (last access: November 2026)
[21] Preliminary tests with a few members of the authors' research group confirmed it.

*Table 3. Parts of the video clip prepared for the assessment*

| Parts | Content | Genre | TimeLine | Source |
|---|---|---|---|---|
| 1 | Valencian Mascletà[22] | Action | 00:00 - 02:25 | À Punt TV |
| 2 | News | News | 02:25 - 03:54 | À Punt TV |
| 3 | Sports | Sports | 03:34 - 06:10 | À Punt TV |
| 4 | Dirt 2 | Commercial | 06:10 - 07:02 | [18] |
| 5 | Mikado[23] | Commercial | 07:02 - 07:18 | YouTube |
| 6 | Fruit Salad[24] | Commercial | 07:18 - 08:37 | YouTube |
| 7 | Nespresso[25] | Commercial | 08:37 - 09:03 | YouTube |
| 8 | The Rustik Bakery[26] | Commercial | 09:03 - 09:24 | YouTube |
| 9 | Mirada Salvatge | Documentary | 09:24 - 11:34 | À Punt TV |
| 10 | Rambo 4 | Movie | 11:44 - 12:40 | [18] |
| 11 | Everest[27] | Movie | 12:40 - 15:11 | YouTube |
| 12 | Avatar[28] | Movie | 15:11 - 17:40 | YouTube |

## 5.2 Participants

A total of 108 non-expert users participated in the subjective tests. Two of them were excluded from the data analysis due to incoherent responses (e.g., contradictory answers). Consequently, the final sample size for the evaluation consisted of 106 valid responses. The final group consisted of 53 females and 53 males (50%-50%), all over the age of 16, with no reported sensory problems relevant to the assessment. The age distribution is shown in Table 4. The sample's average age was approximately 31.35 years (standard deviation: ±15.3).

*Table 4. Participants' age distribution*

| Range (years) | Total | | Men | | Women | |
|---|---|---|---|---|---|---|
| | Number | Percentage | Number | Percentage | Number | Percentage |
| **16-25** | 64 | 60.38% | 35 | 66.04% | 29 | 54.72% |
| **26-35** | 10 | 9.43% | 6 | 11.32% | 4 | 7.55% |
| **36-45** | 4 | 3.77% | 2 | 3.77% | 2 | 3.77% |
| **46-55** | 17 | 16.04% | 4 | 7.55% | 13 | 24.53% |
| **56-65** | 9 | 8.49% | 5 | 9.43% | 4 | 7.55% |
| **>66** | 2 | 1.89% | 1 | 1.89% | 1 | 1.89% |
| **Total** | 106 | 100% | 53 | 100% | 53 | 100% |

The participants' profiles showed considerable diversity. They were students, researchers, teachers, and service workers (administrative staff, security guards, waiters/waitresses, maintenance team members, etc.) recruited from our University Campus, as well as some of the authors' relatives. As the tests were conducted within a university environment, most of the participants held a university degree (55.66%), and a significant proportion held higher-level qualifications, such as postgraduate degrees (8.49%) or doctoral degrees (12.26%). The study also included participants with other

[22] Daytime firework display, which has deep-rooted cultural importance in the Valencian Community.
[23] Mikado: https://youtu.be/bgrmqotNKh (last access: November 2026)
[24] Fruit Salad: https://youtu.be/_IZD_dX9T1Y?list=PLGkRdL5cclh9mO5DRpJnqdk8iiygBVWs_ (last access: November 2026)
[25] Nespresso: https://youtu.be/OS0zmKB2uQs (last access: November 2026)
[26] The Rustik Bakery: https://youtu.be/ZPV9I9OYC9k (last access: November 2026)
[27] Everest: https://youtu.be/YUpnY2W1clw (last access: November 2026)
[28] Avatar: https://youtu.be/bf38f_JINyw (last access: November 2026)

educational backgrounds, such as secondary education (18.87%) or primary education (4.72%).

At the time of the assessment, 39.62% of participants were employed, 40.57% were students, and 16.04% were both studying and employed. The sample also included unemployed individuals (0.94%), homemakers (0.94%), and retirees (1.89%).

Regarding prior experience with similar systems, only 25.47% of participants had experienced mulsemedia systems. Among them, the majority had found such systems in amusement parks, although isolated cases involving 4D cinemas or video game conventions were also reported. Table 5 shows the relationship between participants' educational backgrounds and their prior exposure to mulsemedia technologies. A Chi-square test confirms that participants with a university degree had a significantly higher rate of prior experience than those without (30.9% versus 8.0%, $p = 0.022$).

*Table 5. Participants' prior experience with mulsemedia systems versus their educational level.*

| Participants | Without prior experience | | | With prior experience | | |
|---|---|---|---|---|---|---|
| | Total | Men | Women | Total | Men | Women |
| **Without a university degree (n=25)** | 23 (92.0%) | 12 (48.0%) | 11 (44.0%) | 2 (8.0%) | 1 (4.0%) | 1 (4.0%) |
| **With a university degree (n=81)** | 56 (69.1%) | 27 (33.3%) | 29 (35.8%) | 25 (30.9%) | 13 (16.0%) | 12 (14.8%) |
| **Total (n=106)** | 79 (74.5%) | 39 (36.8) | 40 (37.7%) | 27 (25.5%) | 14 (13.2%) | 13 (12.3%) |

## 5.3 Evaluation Methodology

A total of 53 valid assessment sessions were conducted, each with a couple of participants. Both were seated on the two-seater sofa in front of the TV set in the described mulsemedia scenario, with the sensory effect generation devices surrounding them.

Since participants were already accustomed to watching traditional TV in their daily lives, it was not considered necessary to first show them the video without effects. They could naturally compare this new experience against their everyday habits without needing to watch the content twice. Moreover, repeating the video would have made the test too long, potentially causing fatigue among participants and making recruitment harder.

The evaluation process was divided into 3 stages. In the first one, the participants received a brief explanation of the study they would participate in, along with an overview of the mulsemedia scenario (the existence and location of the sensory effect generation devices in it), the effects they would feel, and how to correctly wear the haptic vest. Then, they signed an informed consent form, agreeing to participate in the assessment. In the second stage, the participants watched the entire video clip on TV and experienced the specified sensory effects at the specified intervals (see Appendix I). The haptic vest was initially worn by one participant at the beginning of this stage, so the broadcast was paused halfway through the session, and it was transferred to the second participant to ensure both users experienced both types of haptic feedback: the haptic vest's vibration and the sofa's vibration. In the third stage, the participants completed a short questionnaire divided into several sections: demographic questions, questions about previous related experiences, questions about the mulsemedia experience (QoE improvement, enjoyment, sync of sensory effects, appropriateness of the included sensory effects, feeling of realism and immersion, distraction, etc.), about their opinion regarding the future adoption of multisensory systems, and about possible improvements. Many of the questions used 5-level Likert scales, and the MOS (Mean Opinion Score) metric was calculated. The results obtained are summarized and discussed in the following sections.

Regarding the estimated duration of each stage, approximately, the initial briefing and preparation phase lasted 5 minutes, the second phase, involving video clip playout, lasted 20 minutes, and finally, completing the questionnaire took between 5 and 10 minutes, depending on the participant's reading speed and whether they required clarification regarding the mulsemedia system or the questionnaire itself. In total, each evaluation session took less than 35 minutes.

## 5.4 Results

This section summarises the evaluation results, divided into several subsections.

### 5.4.1 Results regarding the mulsemedia experience

This subsection details the results obtained regarding the multisensory TV experience, divided into several key areas: the overall enjoyment and participants' QoE improvement, their perception of sensory effect sync with the video content, the appropriateness of the sensory effects included in the assessment and their acceptance by the participants, the level of immersion during the experience, the distraction the sensory effects activation may cause during the TV consumption experience, and, finally, the participants' opinion toward the adoption of multisensory systems in the real world.

#### *5.4.1.1 Enjoyment and improved QoE when sensory effects are included*

Participants were asked to rate their experience during the consumption of the TV with sensory effects by using a 5-level Likert-type scale, with the following levels: 1-'Bad', 2-'Deficient', 3-'Neutral', 4-'Good', and 5-'Excellent'. The combination of TV plus sensory effects provided a highly satisfactory experience for most users. Nearly 92% of the participants rated the experience favorably (59.43% as 'Good' and 32.08% as 'Excellent'). The rest (8.49%) reported it as 'Neutral'. The MOS value was 4.24, with a standard deviation (St. Dev.) of 0.59 (4.21 for men and 4.26 for women, with no statistically significant difference between their responses: p=0.695 in the Mann-Whitney U test). Additionally, the results were analyzed regarding participants' age and educational level using the Kruskal-Wallis and Mann-Whitney U tests, revealing no statistically significant differences.

Participants also rated their agreement with the statement '*I enjoyed the TV + mulsemedia experience*' by using another 5-level Likert scale with the following levels: 1-'Totally Disagree', 2-'Partially Disagree', 3-'Neutral', 4-'Partially Agree' and 5-'Totally Agree'. The results reveal a highly positive reception. 99.05% of the participants agreed with the statement (75.47% as 'Totally Agree', and 23.58% as 'Partially Agree'). A minority of 0.94% reported it as 'Neutral', while there was absolutely no disagreement (0%). The MOS value was 4.75, with a St. Dev. of 0.46 (4.74 for men and 4.75 for women, with no statistically significant difference between their responses: p=0.959 in the Mann-Whitney U test). The statistical analysis (Kruskal-Wallis test) of the responses across participants' age groups revealed no statistically significant differences. Regarding the relationship between participants' educational background and their overall enjoyment of the multisensory hybrid TV experience, the 25 participants without a university degree reported a mean enjoyment score of 4.92. In contrast, the 81 participants with a university degree reported a slightly lower mean score of 4.69. A Mann-Whitney U test confirms that this difference is statistically significant (p = 0.029). Participants who had previously experienced mulsemedia systems reported lower enjoyment scores than first-time users (4.56 versus 4.81, p = 0.007 in the Mann-Whitney U test). The novelty factor clearly affects the participants' QoE ratings. University-educated individuals have a significantly higher rate of prior experience with these technologies than the non-university group (30.9% versus 8.0%, p = 0.022 in the Mann-Whitney U test). The lower scores observed among the University-educated participants can be attributed to prior exposure to technology.

The participants were also asked about their level of agreement with the statement '*Multisensory effects enhance the TV consumption experience*' by using the same scale. 95.28% of the participants agreed with the statement (61.32% as 'Totally Agree', and 33.96% as 'Partially Agree'). A minority of 2.83% reported it as 'Neutral', while disagreement was negligible (1.88% and 0.94% reported it as 'Partially Disagree' and 0.94% as 'Totally Disagree', respectively). The MOS value was 4.54 with a St. Dev. of 0.69 (4.42 for men and 4.66 for women, with no statistically significant differences between their responses: p=0.124 in the Mann-Whitney U test). The statistical analysis (Kruskal-Wallis test) of the responses across participants' age groups revealed no statistically significant differences. Nevertheless, despite the overwhelming consensus that multisensory content significantly enhances the TV viewing experience, the perceived enhancement varies by educational background, with a statistically significant difference (p = 0.002 in the Mann-Whitney U test). Non-university participants agreed more strongly with the statement (4.88) than university-educated participants (4.43). While the technology is widely appreciated, higher-profile academics tend to take a slightly more conservative, critical stance on its overall impact on traditional TV consumption.

A key indicator of multisensory hybrid TV services is whether the sensory effects add value to the TV-consumption experience. A user might not personally like a specific sensory effect (e.g., for subjective reasons) but still recognize that it enhances the experience. So, participants were also asked to indicate their level of agreement with the statement: *'The sensory effects added value to the audiovisual content viewed on TV'.* 96.22% agreed (72.64% 'Totally' and 23.58% 'Partially'), 2.83% were 'Neutral', and only 0.94% 'Partially Disagreed' with the statement. The MOS value was 4.68 with a St. Dev. of 0.58 (4.57 for men and 4.79 for women, with a statistically significant difference between their responses: p=0.047 in the Mann-Whitney U test). Women perceived the sensory effects as adding significantly more value to the content than men did. The statistical analysis (Kruskal-Wallis test) of the responses across participants' age groups did not reveal statistically significant differences. Nevertheless, the analysis of the data reveals that educational background also modulates this perception (p = 0.045 in the Mann-Whitney U test). Non-university individuals rated the added value significantly higher (4.88) than university-educated participants (4.62).

Regarding the perceived realism of the mulsemedia consumption experience compared to a traditional TV consumption experience (i.e., without sensory effects), participants were asked to answer the question: *'How realistic did you feel the multisensory hybrid TV experience compared to the traditional TV (i.e., without sensory effects that stimulate other senses beyond sight and hearing)?'* by using another 5-level Likert-type scale with the following levels: *1-'Much Less Realistic'*, *2-'A Bit Less Realistic'*, *3-'Equally Realistic'*, *4-'A Bit More Realistic'* and *5-'Much More Realistic'*. Most participants (95.28%) found the mulsemedia experience to be more realistic (49.06% 'A bit more' and 46.23% 'Much more') than traditional TV viewing. The rest (4.72%) rated it as 'Equally realistic', while no participant rated it as 'Less realistic'. The MOS value was 4.42, with a St. Dev. of 0.58 (4.40 for men and 4.42 for women, with no statistically significant difference between their responses: p=0.799 in the Mann-Whitney U test). The statistical analysis of responses across participants' age and educational background groups (Kruskal-Wallis and Mann-Whitney U tests, respectively) revealed no statistically significant differences.

Users also rated their level of agreement regarding the sentence '*I would repeat this type of experience'*. 94.34% of the participants agreed with the statement (85.85% as 'Totally Agree', and 8.49% as 'Partially Agree'). Most users genuinely enjoyed the experience and are willing to repeat it. A small fraction of participants remained undecided (3.77% rated it as 'Neutral'), while two participants (1.88%) expressed disagreement (1 as 'Partially Disagree' and 1 as 'Totally Disagree'). The MOS value was 4.77 with a St. Dev. of 0.65 (4.75 for men and 4.79 for women, with no statistically significant difference between their responses: p=0.822 in the Mann-Whitney U test).

Participants were asked whether they would recommend this experience to others. They provided their level of agreement with the statement '*I would recommend this experience to others'.* Although, as shown above, a few participants indicated they did not enjoy the experience or considered that mulsemedia effects do not improve the traditional TV consumption experience, nearly all participants would recommend this experience to others in their environment. 97.17% of the participants agreed (79.25% 'Totally', and 17.92% 'Partially') with the statement. The rest (2.83%) rated it as 'Neutral'. The MOS value was 4.76 with a St. Dev. of 0.49 (4.77 for men and 4.75 for women, with no statistically significant difference between their responses: p=0.683 in the Mann-Whitney U test). The statistical analysis identified a significant difference across the participants' age groups (p = 0.049 in the Kruskal-Wallis test). The highest levels of advocacy were found among younger participants aged 16-25, who reported a MOS value of 4.81, and middle-aged adults aged 46-55, who reported a MOS value of 4.88. Conversely, the 56-65 age group adopted a more conservative stance, registering the lowest recommendation score in the study, with a mean of 4.33. The two senior participants (>66 years old) would recommend the experience (MOS=5). The statistical analysis of responses across participants' educational background groups (Mann-Whitney U tests, respectively) did not reveal statistically significant differences.

Participants were also asked to indicate their level of agreement with the statement: *'I think I would like to use this TV + mulsemedia scenario frequently.'* 62.27% of participants agreed (29.25% 'Totally' and 33.02% 'Partially'). However, this statement elicited the highest level of hesitation in the study: 17.92% of the participants rated it as 'Neutral', and 19.81% disagreed (10.38% 'Partially', and 9.43% 'Totally'). The MOS value was 3.63 with a St. Dev. of 1.27 (3.51 for men and 3.74 for women, with no statistically significant difference between their responses: p=0.353 in the Mann-Whitney U test). The statistical analysis of responses across participants' age and educational background groups (Kruskal-Wallis and Mann-Whitney U tests, respectively) revealed no statistically significant differences.

#### *5.4.1.2 Level of immersion*

Participants were also asked to evaluate their perceived level of immersion by providing their level of agreement with the statement: '*The experience was immersive, that is, the sensory effects increased my feeling of immersion, and I felt more integrated/submerged during the TV consumption session*'. The results indicate strong consensus, with 94.34% of participants agreeing (57.55% 'Totally' and 36.79% 'Partially') that the mulsemedia setup enhanced their immersion. 3.77% were 'Neutral', and only 1.89% 'Partially Disagreed'. The overall MOS value achieved for this metric was 4.50 with a St. Dev. of 0.67. Females reported a slightly higher level of immersion than males (a MOS of 4.57 versus 4.43, respectively); however, a Mann-Whitney U test showed no statistically significant difference between their responses. The statistical analysis of responses across participants' age and educational background groups (Kruskal-Wallis and Mann-Whitney U tests, respectively) revealed no statistically significant differences.

#### *5.4.1.3 Synchronization of sensory effects with TV content playout*

To assess the proper functioning of the implemented prototype, the perceived synchronization between sensory effects and the audiovisual content presented on the TV was also evaluated. Participants were asked to indicate their level of agreement with the statement: *'During the consumption of the audiovisual content on the TV, I felt that the sensory effects were synchronized with it.'* The results were overwhelmingly positive. 95.29% of the participants agreed with the statement (46.23% 'Totally' and 49.06% 'Partially'). In contrast, only 1.89% rated it as 'Neutral', while a mere 2.83% 'Partially' disagreed with the statement. The MOS value was 4.39 with a St. Dev. of 0.67 (4.34 for men and 4.43 for women, with no statistically significant difference between their responses: p=0.324 in the Mann-Whitney U test). The statistical analysis (Kruskal-Wallis test) of responses across participants' age did not reveal statistically significant differences. Although overall sync perception was highly rated by participants, a Mann-Whitney U test revealed statistically significant differences by participants' educational background (p = 0.018). Non-university individuals reported a higher perceived sync score (4.64) compared to university-educated participants (4.31). This difference reinforces the observed pattern where higher academic profiles evaluate the technical performance of the mulsemedia integration with a more critical and demanding approach.

#### *5.4.1.4 Sensory effects appropriateness*

To assess the appropriateness of the chosen effects included in the experience, the following aspects were considered:

- *Consistency of Sensory Effects with AV media content:* 93.4% of participants agreed (55.66% 'Totally' and 37.74% 'Partially') with the statement '*The sensory effects experienced were consistent with the audiovisual content'*. In contrast, only 2.83% were 'Neutral', while 3.77% 'Partially' disagreed. The MOS value was 4.45 with a St. Dev. of 0.73 (4.47 for men and 4.43 for women, with no statistically significant difference between their responses: p=0.903 in the Mann-Whitney U test). These results indicate a strong participants' validation of the specific effects chosen by the authors for the selected video clip, confirming that participants perceived their occurrence as logical and meaningful within the narrative context. The statistical analysis (Kruskal-Wallis test) of responses across participants' age groups did not reveal statistically significant differences. Although the overall validation of the mulsemedia content's coherence was strong, consistently with previous findings on distribution and relevance, perceived coherence is also significantly influenced by the user's educational background. The statistical analysis (p = 0.001 in the Mann-Whitney U test) confirmed that non-university participants rated this consistency significantly higher (4.84) than university-educated participants (4.33).

- *Balance and distribution of the sensory effects:* Participants were asked about their level of agreement with the statement '*The sensory effects were well distributed and balanced during the TV video sequence'*. 90.57% agreed (52.83% 'Totally' and 37.74% 'Partially'). In contrast, 8.49% were 'Neutral', while only 0.94% 'Partially' disagreed. The MOS value was 4.42, with a St. Dev. of 0.69 (4.32 for men and 4.53 for women, with no statistically significant differences between their responses: p=0.087 in the Mann-Whitney U test). This suggests successful execution of the video design and effect assignment. The statistical analysis (Kruskal-Wallis test) of responses across participants' age groups did not reveal statistically significant differences. Despite the high ratings, a statistical analysis reveals a highly significant difference across participants' educational backgrounds. A Mann-Whitney U test confirms (p = 0.001) that non-university users rated the balance and distribution of the sensory stimuli significantly higher (4.80) than university-educated participants (4.31).

  Participants also provided their level of agreement with the statement *'I believe that some parts of the audiovisual content displayed on the TV had an excessive number of sensory effects'.* 22.64% agreed (9.43% 'Totally' and 13.21% 'Partially'), 16.04% were 'Neutral', and 61.32% disagreed (37.74% 'Totally' and 23.58% 'Partially') with the statement. Although a majority did not find the effects excessive, the percentage of participants who agreed with the statement is slightly higher than expected. This can be understood given the test's compact nature, which was designed to allow participants to experience the maximum number of sensory effects in the shortest possible time, without extending the evaluation session. This feedback will help the authors to calibrate the saturation levels of sensory effects better for future-designed mulsemedia experiences. Results about mental fatigue, which can be directly related to this issue, are presented later. The MOS value was 2.33 with a St. Dev. of 1.91 (2.34 for men and 2.32 for women, with no statistically significant difference between their responses: p=0.966 in the Mann-Whitney U test). The statistical analysis (Kruskal-Wallis test) of responses across participants' age groups did not reveal statistically significant differences. Nevertheless, participants' perceptions of sensory overload are also influenced by their educational backgrounds ($p = 0.048$ in the Mann-Whitney U test). University-educated participants were more likely to perceive an excessive number of effects (2.47) than non-university individuals (1.88). This heightened sensitivity to sensory saturation among higher academic profiles aligns with their generally more critical evaluation of the system, further reinforcing the recommendation to implement customizable sensory activation and intensity profiles to accommodate varying user tolerances.

- *Relevance of the selected effects:* Participants were asked to indicate their level of agreement with the statement: *'The sensory effects included in the audiovisual content viewed on the TV were the most relevant for it'.* Most participants (93.4) agreed with the statement (57.55% 'Totally' and 35.85% 'Partially'), while the rest were 'Neutral'. No disagreement was reported, indicating that the selected effects matched the content of the prepared video clip well and were suitable for the context. The MOS value was 4.51 with a St. Dev. of 0.62 (4.51 for men and 4.51 for women, with no statistically significant difference between their responses: $p=0.772$ in the Mann-Whitney U test). The statistical analysis (Kruskal-Wallis test) of responses across participants' age groups did not reveal statistically significant differences. Even though there was no disagreement with the statement, an additional statistical analysis of the responses reveals a significant disparity in participants' educational backgrounds. A Mann-Whitney U test ($p = 0.007$) shows that non-university individuals rated the contextual relevance of the effects significantly higher (4.80) than university-educated participants (4.42).

- *Intensity and realism of sensory effects:* It was also analyzed whether each of the sensory effects included in the experience, generated from different locations around the user, were perceived by users with sufficient intensity and whether they found their experience realistic. Another 5-level Likert scale was used, with the following five levels: *1-'Bad', 2-'Poor', 3-'Fair', 4-'Good',* and *5-'Excellent'*. Figure 19 and Table 6 summarize the results obtained for all the experienced effects. Overall, the perceived realism of the effects was rated positively (with almost all MOS values > 4), featuring a high proportion of responses in the 'Excellent' and 'Good' categories. Notably, these results closely align with findings on the participants' level of immersion. When users perceive the TV experience as more realistic, they feel more immersed in it. The most highly rated effects were ambient light and wind (both tied for the highest MOS), followed by cooling, scents, haptic vest vibration, heating, and sofa vibration. Men and women rated the realism of the effects similarly (no statistically significant differences were found), although females generally provided slightly higher ratings across most stimuli (particularly wind and cooling), except for scent effects, which males rated higher (4.23 compared to 4.00). Among males, regarding the positive ('Excellent' or 'Good') responses, the 'Excellent' responses were predominant for ambient light, scents, and haptic vest vibration. There were a moderate number of responses rated as 'Fair' across all effects, some responses rated as 'Poor' (most notably, 7 for cooling), and very few responses rated as 'Bad' (2 for sofa vibration and 1 for haptic vest vibration). In contrast, among females, although most responses were positive, those rated 'Excellent' predominated for wind, haptic vest vibration, cooling, ambient light, and heating, compared with those rated ‘Good’. Notably, there were no ‘Poor’ or ‘Bad’ ratings from females for ambient light, wind, or cooling. However, a few responses were rated ‘Poor’ (4 for sofa vibration) and ‘Bad’ (2 for heating, 1 for scents, and 1 for haptic vest vibration). This indicates that while visual and wind effects were universally perceived as highly realistic, the thermal, scent, and haptic effects presented slightly more variability among users.

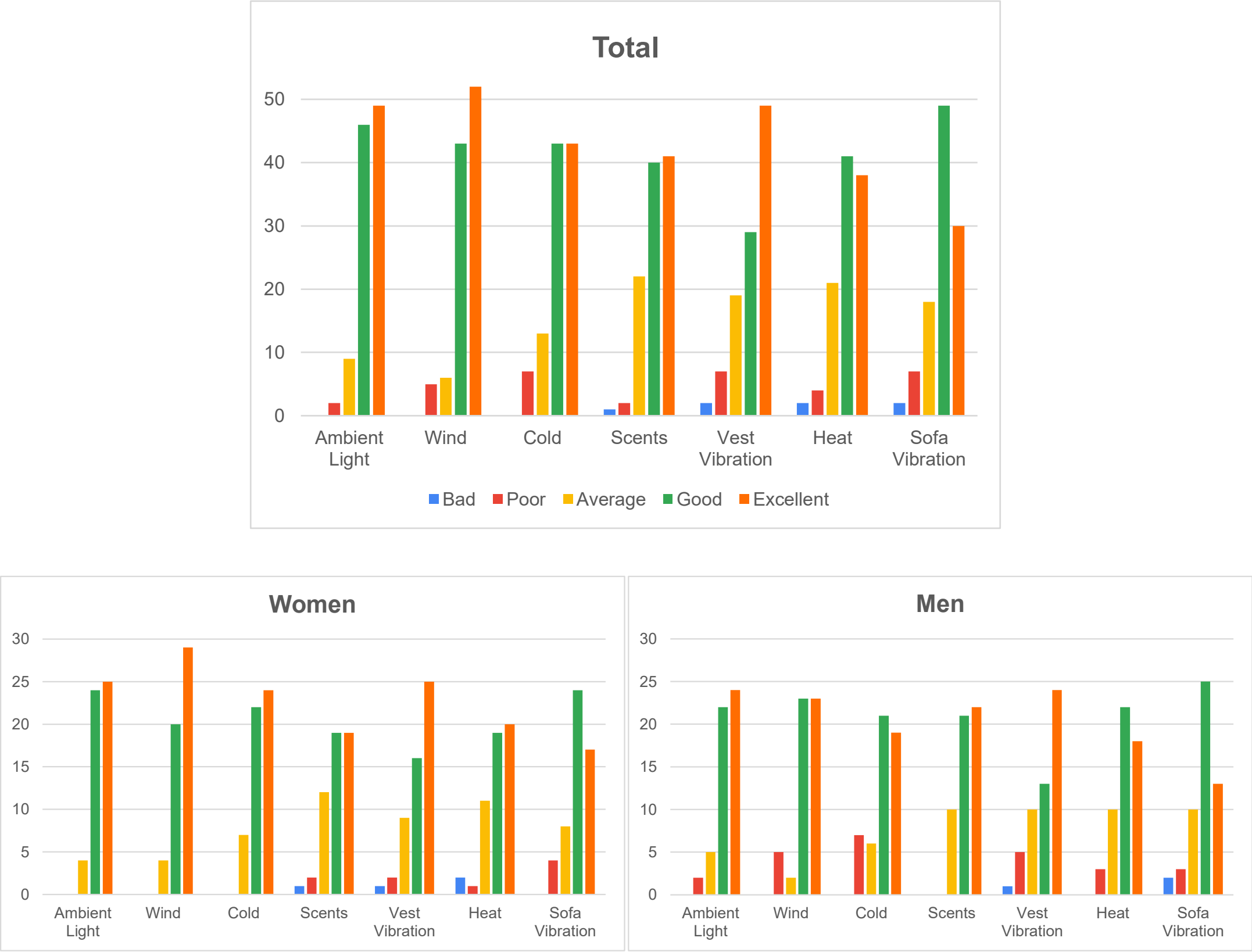

*Figure 19: Perception of the realism of sensory effects during the experience*

*Table 6. MOS values and the standard deviation of the perception of the realism of sensory effects during the experience*

| Sensory Effects | Total | | Males | | Females | |
|---|---|---|---|---|---|---|
| | MOS | St. Dev. | MOS | St. Dev. | MOS | St. Dev. |
| **Ambiental light** | 4,34 | 0,72 | 4,28 | 0,79 | 4,40 | 0,63 |
| **Wind** | 4,34 | 0,79 | 4,21 | 0,91 | 4,47 | 0,64 |
| **Cooling** | 4,15 | 0,88 | 3,98 | 1,01 | 4,32 | 0,70 |
| **Scent** | 4,11 | 0,87 | 4,23 | 0,75 | 4,00 | 0,96 |
| **Haptic Vest vibration** | 4,09 | 1,04 | 4,02 | 1,10 | 4,17 | 0,98 |
| **Heating** | 4,03 | 0,94 | 4,04 | 0,88 | 4,02 | 1,01 |
| **Sofa vibration** | 3,92 | 0,94 | 3,83 | 0,99 | 4,02 | 0,89 |

To provide a more detailed analysis, the MOS values for perceived realism were broken down by individual sensory effect and gender. Ambient light (4.28 men, 4.40 women) and wind (4.21 men, 4.47 women) received some of the highest realism ratings. Across all evaluated stimuli, including sofa vibration, vest vibration, aromas, cooling, and heating, women consistently reported slightly higher or comparable MOS values to men, with the largest divergence observed in the perception of cold. The Mann-Whitney U test confirmed that there were no statistically significant differences between men's and women's responses for any of the individual sensory effects ($p > 0.05$ in all cases). The results were analyzed across participants' age groups using the Kruskal-Wallis test, and no statistically significant differences were found. However, significant disparities were also observed among participants with different educational backgrounds (Table 7). According to a Mann-Whitney U test, non-university participants rated the realism of certain sensory effects significantly higher than university-educated participants did. Specifically, the realism of the sofa vibration was rated at 4.32 by the non-university group, compared with 3.80 by the university group ($p = 0.018$ in the Mann-Whitney U test). The most substantial difference was observed in the perception of aromas, which achieved a 4.56, a highly realistic score among non-university participants, compared with 3.98, a more moderate rating among university participants ($p = 0.002$). Similarly, the heating effect followed the same trend (4.44 versus 3.90; $p = 0.016$). These results reinforce the pattern that higher academic profiles tend to scrutinize sensory stimuli more strictly, demanding a higher degree of physical fidelity to perceive the simulated effects as genuinely realistic, whereas non-university profiles demonstrate a much higher baseline of sensory acceptance and immersion.

*Table 7. MOS values and standard deviation of the Perceived Realism of Multisensory Effects by Educational Level*

| Sensory Effect | Non-University (n=25) | | University student (n=81) | |
|---|---|---|---|---|
| | MOS | St. Dev. | MOS | St. Dev. |
| **Sofa vibration** | 4.32 | 0.69 | 3.80 | 0,98 |
| **Aromas** | 4.56 | 065 | 3.98 | 0,88 |
| **Heating** | 4.44 | 0.58 | 3.90 | 1,00 |
| **Ambient light** | 4.32 | 0.80 | 4.35 | 0,69 |
| **Vest vibration** | 4.40 | 0.82 | 4.00 | 1,08 |
| **Wind** | 4.56 | 0.58 | 4.27 | 0,84 |
| **Cooling** | 4.44 | 0.71 | 4.06 | 0,91 |

- *Fatigue during the test:* A total of 13 participants (12%, 7 females and 6 males) reported experiencing fatigue during the assessment. Based on their qualitative feedback, this sensory overload could be attributed to four main factors. First, the high intensity and prolonged duration of the haptic vest vibration caused some of them discomfort or dizziness, particularly during action-heavy sequences. Second, the olfactory stimuli led some others to saturation when scents overlapped, were delivered too rapidly in succession (e.g., multiple fruit aromas), or lingered excessively in the room (e.g., bread or ash). Third, others emphasized that the frenetic pace and rapid transitions of certain audiovisual segments, such as trailers, commercials, and action scenes, contributed to cognitive overload. Finally, some participants experienced a general multimodal saturation due to a lack of familiarity with simultaneous sensory stimuli (such as the combination of lights, scents, and haptic feedback) coupled with abrupt content changes. An interesting finding emerged from analyzing participants' mental fatigue during the evaluation across age groups. The data revealed a statistically significant generational divide in responses to sensory overload (Fisher's exact test, $p = 0.013$). Specifically, all participants who reported feeling mental fatigue belonged to the younger demographic (under 35 years old), representing 17.6% of that specific age group (13 out of 74 users). In stark contrast, absolutely none of the 32 participants aged 35 or older reported experiencing any mental fatigue (0%). This result may suggest that younger participants were more susceptible to mental fatigue in high-density mulsemedia environments than older participants. Additionally, the educational level variable was analyzed using the Mann-Whitney U test, revealing no statistically significant differences.

### *5.4.1.5 Acceptance of the sensory effects*

Although the general acceptance of the multisensory hybrid TV experience was already established, a more granular analysis of the acceptance of each sensory effect was conducted. The level of acceptance of the sensory effects by the participants was evaluated using another 5-level Likert scale with the following levels: 1-*'I did not like it at all' (total disapproval), 2-'I did not like it' (dislike), 3-'Neutral', 4-' I liked it' (like),* and *5- 'I liked it very much' (total acceptance)*. Figure 20 and Table 8 summarize the results obtained for all the experienced effects.

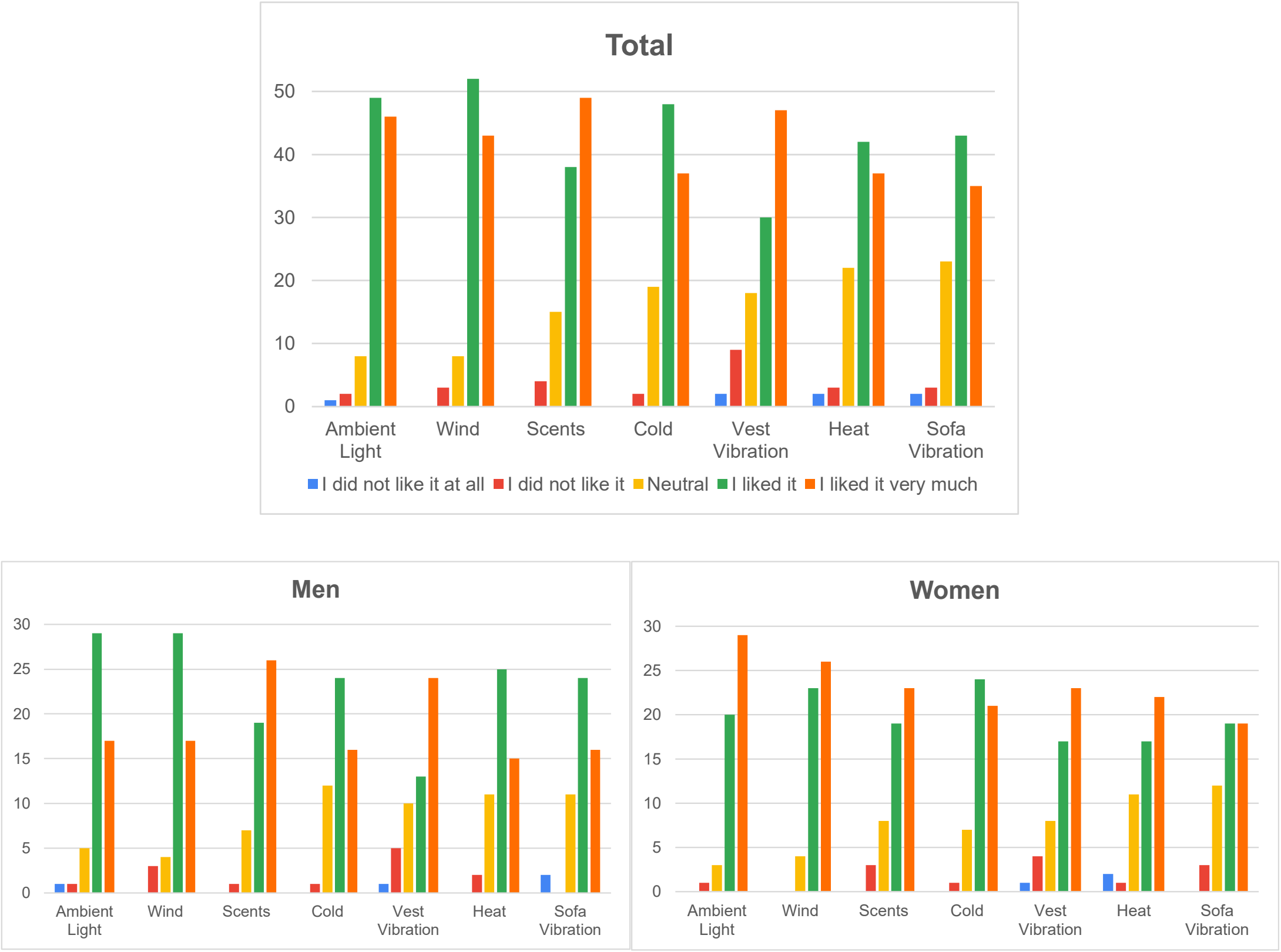

*Figure 20. Level of acceptance of the experienced sensory effects*

*Table 8. MOS values and the standard deviation of the acceptance of the experienced sensory effects*

| **Sensory Effects** | **Total** | | **Males** | | **Females** | |
|---|---|---|---|---|---|---|
| | **MOS** | **St. Dev.** | **MOS** | **St. Dev.** | **MOS** | **St. Dev.** |
| Ambient light | 4.29 | 0,77 | 4.13 | 0,81 | 4.45 | 0,70 |
| Wind | 4.28 | 0,72 | 4.13 | 0,79 | 4.42 | 0,63 |
| Scent | 4.25 | 0,84 | 4.32 | 0,78 | 4.17 | 0,89 |
| Cooling | 4.14 | 0,77 | 4.04 | 0,78 | 4.23 | 0,75 |
| Vest vibration | 4.05 | 1,06 | 4.02 | 1,10 | 4.08 | 1,03 |
| Heating | 4.03 | 0,92 | 4.00 | 0,81 | 4.06 | 1,03 |
| Sofa vibration | 4 | 0,92 | 3.98 | 0,93 | 4.02 | 0,91 |

Regarding qualitative feedback on disliked effects from participants, the haptic vest received the most comments, but only from 4 participants (2 females and 2 males). They stated that the vest's vibration was too intense or overwhelming. Additionally, 3 participants (2 females and 1 male) found the physical requirement of wearing the vest uncomfortable or felt it added no value to the overall experience. Furthermore, 3 males noted a lack of realism, suggesting that the haptic feedback should be more spatially localized to better match the on-screen action. As for the sofa vibration, 3 participants (2 females and 1 male) criticized its lack of intensity, noting it was barely perceptible. Two males also mentioned poor spatial distribution, noting that the vibration felt too focused on the center of the sofa. Nevertheless, the sofa has two subwoofers, one on each side. The remaining sensory effects also received very few criticisms. Two participants (1 female and 1 male) found the thermal effects too weak, and one female specifically disliked the abruptness of the temperature changes. Regarding scents, 1 female requested a higher intensity, while another stated that '*she found some strong, brief scents highly distracting*'. One male considered ambient light distracting from the TV screen, and 1 female suggested that the lighting transitions should be more radical. One male stated that the wind effect failed to increase his immersion.

Regarding participants' acceptance of sensory effects, the ambient light effect was the most accepted, followed by wind, scents, cooling, vest vibration, heating, and sofa vibration. The results indicate a clear distinction between ambient effects and more direct stimuli. Ambient light and wind were perceived as softer, less intrusive enhancements (i.e., immersive elements that accompany the experience without demanding the user's primary focus). Consequently, these effects had the lowest rejection rates, with fewer participants reporting that they 'did not like at all'. For the ambient light effect, the MOS value was 4.29 (4.13 for men and 4.45 for women). Wind also achieved a high MOS value of 4.28 (4.13 for men and 4.42 for women). In contrast, more prominent effects, such as scents (MOS value of 4.25; 4.32 for men and 4.17 for women) and, most notably, vibration (MOS value of 4.05; 4.02 for men and 4.08 for women), elicited more polarized responses. The haptic vest presents a compelling case: while it ranked second in the 'liked very much' category, it also received the highest rate of strong disapproval. The Mann-Whitney U test showed a statistically significant difference between men's and women's responses regarding acceptance of ambient light ($p=0.021$), with women rating this effect significantly higher. For all other sensory effects, no statistically significant differences were observed between genders ($p > 0.05$). The results were also analyzed across participants' age groups using the Kruskal-Wallis test, revealing no statistically significant differences. Nevertheless, significant variations in the acceptance of specific sensory effects by participants' educational background were also observed (

Table 9). A Mann-Whitney U test revealed that non-university participants reported a significantly higher acceptance of the scent effect than university-educated participants (4.64 versus 4.12, $p = 0.010$). For the remaining effects, no statistically significant differences were found.

*Table 9. MOS values and the standard deviation of the acceptance of the experienced sensory effects across participants' educational levels*

| Sensory Effect | Non-University (n=25) | | University student (n=81) | |
|---|---|---|---|---|
| | MOS | St. Dev. | MOS | St. Dev. |
| Ambient light | 4.32 | 0,69 | 4.28 | 0,98 |
| Wind | 4.48 | 0,65 | 4.21 | 0,88 |
| Scent | 4.64 | 0,58 | 4.12 | 1,00 |
| Cooling | 4.40 | 0,80 | 4.05 | 0,69 |
| Vest vibration | 4.24 | 0,82 | 3.93 | 1,08 |
| Heating | 4.28 | 0,58 | 3.95 | 0,84 |
| Sofa vibration | 4.32 | 0,71 | 3.96 | 0,91 |

To compare the two types of vibration effects available in the TV mulsemedia scenario (sofa and haptic vest), participants were asked which one they preferred. The *haptic vest* vibration effect was preferred by 47.17% of participants (52.83% of men and 41.51% of women), whereas 24.53% preferred the sofa vibration effect (22.64% of men and 26.42% of women). An identical percentage, 24.53% (22.64% of men and 26.42% of women), showed no preference between the two vibration types. The rest (3.77%, 1.89% of men and 5.66% of women) did not like any of the vibration effects. There were no statistically significant differences between men's and women's responses (p=0.571, in the Chi-square test). Despite a preference for the haptic vest's vibration effect, qualitative feedback indicated that some users found the device's physical requirement of wearing it cumbersome or uncomfortable, diminishing its appeal despite its superior vibration intensity. An interesting phenomenon was also observed: several participants reported that while wearing the vest, they were unaware that the sofa was also vibrating. It was only after removing the vest that they perceived the sofa's feedback. This suggests a sensory masking effect, in which the localized, intense stimulus from the vest overrode the more subtle, distributed vibration of the sofa.

All responses (including individual scale ratings and preferences regarding vibration effects) were statistically analyzed by participants' age and educational background groups (Kruskal-Wallis and Mann-Whitney U tests, respectively), and no statistically significant differences were found.

#### 5.4.1.6 Distraction

The potential distraction from AV content due to perceived sensory effects was also analyzed. Users were asked to indicate their level of agreement with the statement: *'The sensory effects were a distraction or hindrance from the TV experience'*. 61.32% of the participants disagreed (35.85% 'Totally' and 25.47% 'Partially'), 18.87% were 'Neutral', and 19.81% agreed (12.26% 'Partially' and 7.55% 'Totally'). The MOS value was 2.30 with a St. Dev. of 1.9 (2.47 for men and 2.13 for women, with no statistically significant difference in their responses: p=0.112 in the Mann-Whitney U test). Although the overall data confirm that sensory effects were not a major source of distraction for most users, an analysis of individual stimuli reveals a marginal gender difference in the heating effect. In a Mann-Whitney U test (p = 0.049), men reported a slightly higher level of heating-related distraction (1.87) than women (1.66). While both averages remain quite low (reinforcing the general conclusion that the effects were not disruptive), this subtle variation suggests that male participants were slightly more sensitive to thermal stimuli during the multisensory hybrid TV experience. The statistical analysis of responses across participants' age and educational background groups (Kruskal-Wallis and Mann-Whitney U tests, respectively) revealed no statistically significant differences.

#### *5.4.1.7 Adoption of multisensory TV systems*

Having established that most users enjoyed the TV mulsemedia experience and reported generally positive opinions, it was considered pertinent to analyze how participants perceived the adoption of sensory effects in real-world TV-related scenarios. They were asked first to indicate their agreement with the statement: *'If it were possible, I would like to have a TV + mulsemedia system at home'.* 74.53% of them agreed with the statement (51.89% 'Totally' and 22.64% 'Partially'), 18.87% were 'Neutral', and 6.61% disagreed (4.72% 'Partially', and 1.89% 'Totally'). The MOS value was 4.18 with a St. Dev. of 1.02 (4.26 for men and 4.09 for women, with no statistically significant difference between their responses: p=0.695 in the Mann-Whitney U test). The statistical analysis of responses across participants' age and educational background groups (Kruskal-Wallis and Mann-Whitney U tests, respectively) revealed no statistically significant differences.

Although mulsemedia is perceived as a positive addition to the TV experience, a subset of users remained unconvinced regarding its habitual or prolonged use. They perceived the system as suitable for a short demonstration, but potentially overwhelming for a full-length, 3-hour movie. The provided reasons for the disagreement were that they believed the system would be too costly and that they lacked sufficient space or the required infrastructure to install the necessary sensory generation devices at home. The hesitation regarding the daily, prolonged use of these technologies is a universal sentiment. The perceived risk of sensory fatigue during feature-length content was shared across the entire sample, indicating that these practical concerns are inherent to the home TV viewing experience rather than influenced by users' gender, age, or educational background.

This behavioral pattern is further evidenced by the participants' responses regarding their preference involving a hypothetical viewing of a feature film (e.g., Avatar) at home, with or without multisensory effects. They had to answer the question: *'If you had to watch a movie on TV at home (such as, for example, Avatar), how would you prefer to watch it?'* A 2-level scale was used with the options '*Traditional TV*' and '*TV with mulsemedia effects*'. 20.75% preferred the '*Traditional TV*' format, while 79.25% preferred the '*TV with mulsemedia effects' format* (73.58% of men and 84.91% of women). No statistically significant differences were found between men's and women's responses (p=0.344, in the Chi-square test). Additionally, the results were also analyzed across participants' age groups and educational levels using the Kruskal-Wallis and Mann-Whitney U tests, revealing no statistically significant differences. While most participants clearly favor the multisensory option, these results indicate some resistance to frequent, day-to-day consumption, without sounding alarmist.

Participants were also asked, assuming they could acquire (regardless of cost) any of the sensory effects they had experienced during the session for their home, which one they would acquire. Scent effects emerged as the most desired feature (34.91% in total; 43.40% men and 26.42% women), closely followed by ambient light (33.96% in total;

32.08% men and 35.85% women) and vibration (26.42% in total; 20.75% men and 32.08% women). In sharp contrast, interest in wind was minimal (4.72% in total; 3.77% men and 5.66% women), while thermal effects (heating and cooling) garnered no interest at all (0%). There were no statistically significant differences between men's and women's responses ($p=0.440$ in the Chi-square test). The statistical analysis of responses across participants' age and educational background groups (Kruskal-Wallis and Mann-Whitney U tests, respectively) did not reveal statistically significant differences. The preference hierarchy favoring scents, ambient light, and vibration over thermal and wind stimuli is shared among the participants in the assessment. There was a collective consensus on which mulsemedia features offer the most desirable balance of immersion and comfort for the home environment, regardless of the user's gender, age, or educational background.

Finally, the participants' willingness to pay for this technology was also collected. They were asked to provide their level of agreement with the statement '*I wouldn't mind paying a higher price (e.g., a premium subscription) to experience sensory effects while consuming audiovisual content on TV*'. 64.24% of the participants agreed with the statement (21.70% 'Totally' and 42.45% 'Partially'), 20.75% were 'Neutral', and 15.09% disagreed (9.43% 'Totally' and 5.66% 'Partially'). The MOS value was 3.62 (3.43 for men and 3.81 for women, with no statistically significant difference between their responses: $p=0.143$ in the Mann-Whitney U test), and a St. Dev. of 1.16. The statistical analysis of responses across participants' age and educational background groups (Kruskal-Wallis and Mann-Whitney U tests, respectively) did not reveal statistically significant differences. These results suggest that while the experience was rated positively and enjoyable, for a certain segment of participants, the added value does not justify an increase in the cost of the TV service. However, it is important not to overstate this resistance, as the majority still expressed willingness to pay.

### 5.4.2 Suggested improvements

To improve the overall multisensory hybrid TV scenario for future research, participants were asked which aspects they would suggest changing or improving regarding both the video content and the sensory effects. Feedback on the AV component was generally positive; participants found the content adequate for the test, though 3 participants noted minor video-editing details in a few transitions between parts of the video. The production of the 17-minute AV clip was subject to specific research constraints (such as the absolute necessity to condense six different TV genres and multiple sensory stimuli into a brief session) and resource limitations (including budget constraints, limited resources, and reliance on available consumer-grade devices). Consequently, it required prioritizing experimental utility and stimulus density over the seamless narrative coherence typical of professional TV broadcasts.

Regarding multisensory effects, while some participants reported reduced intensity of thermal sensations (heating and cooling), the most significant feedback focused on the haptic vest's vibration. Suggestions included better spatial localization of the vibration. For example, a participant suggested activating only specific parts of the vest to match on-screen impacts (specifically, in part 10 of the video clip, which contains highlights of the action movie Rambo-4; see Table 3). Another participant suggested using dynamic intensity variation during prolonged vibrations.

## 5.5 Summary and discussion of the results

The presented results show that the implemented multisensory hybrid TV service consistently delivered a highly satisfactory and immersive experience for participants, regardless of their gender, age, or educational background. The combination of TV and sensory effects provided an enjoyable, immersive, and realistic experience for most participants (+90%). Nearly all of them (+95%) recognized that the sensory effects added value to the experience. It was perceived as unique and novel, and an experience worth trying. Most of them (+94%) genuinely enjoyed the experience and were willing to repeat it. Nearly all (+97%) reported that they would recommend this type of experience to others. There was a high level of overall satisfaction and willingness to engage with mulsemedia setups, demonstrating a broad appeal and robust acceptance. Even among the few participants who did not personally enjoy the experience, it was perceived as unique and novel, and an experience worth trying.

According to the results obtained, women were a bit more appreciative of the mulsemedia enhancement, as they perceived the sensory effects as adding significantly more value to the content than men did. Participants appear to prioritize effects that provide a high degree of immersion and surprise (such as scents or lights) or established visual enhancements (ambient light), while discarding atmospheric elements (wind or temperature) that might be perceived as less impactful or potentially intrusive in a domestic setting.

Additionally, it was found that the degree of the provided enhancement did not vary by age but varied significantly by educational background. Participants with a higher academic profile maintained a slightly more conservative and critical stance toward the overall impact of sensory effects on traditional TV consumption. They had higher expectations of the technology, whereas individuals with non-university profiles showed greater enthusiasm.

Despite overwhelming consensus that sensory effects significantly enhance the TV viewing experience, a small subset of participants viewed it as appropriate for sporadic or intermittent TV viewing sessions rather than as a continuous habit in their home environment. Although a substantial portion of the sample was indeed willing to adopt the system for frequent home use, which remains a positive signal for market viability, the hesitation to pay a premium for mulsemedia content in TV services reflects a general sentiment. The perceived value-to-cost ratio presents a universal challenge for the future commercialization of multisensory hybrid TV services, requiring further refinement to justify a premium subscription model.

Sensory perception is highly subjective and heavily dependent on individual thresholds. So, mulsemedia systems in general must offer personalization profiles, allowing users to toggle or adjust specific effects according to their individual sensitivities or preferences. Discomfort and fatigue can be partially (or even fully) reduced if users' preferences are taken into account. So, it is important that these systems also include a module to manage them, allowing users to personalize their multisensory hybrid TV experiences automatically or manually, while avoiding undesirable effects or annoying features (e.g., excessively high intensities or unsupportable odors).

Moreover, as confirmed in the assessment, in mulsemedia systems, sensory effects may be distracting for some users. Artificial Intelligence (AI)-based adaptation techniques could mitigate this issue by dynamically adjusting the intensity of these effects based on user behavior throughout the session or content characteristics. As an example, to achieve this, AI-based models could monitor user signals in real time, including gaze direction, interaction patterns, physiological responses, and explicit feedback, to identify signs of cognitive overload or discomfort and automatically increase or decrease the intensity of the sensory effects accordingly. For instance, low user engagement or frequent pauses may indicate a need to reduce haptic or ambient effects. Additionally, content-aware adaptation could be implemented by using AI to classify media genres and assess narrative intensity. More subdued genres, such as documentaries or educational content, may benefit from subtle sensory cues, whereas action-oriented content may accommodate more pronounced effects. Integrating user-centered and content-aware models would enable sensory intensity to be adjusted in a personalized, context-sensitive manner throughout the session, enhancing immersion and reducing distraction without requiring manual user intervention. However, the design and implementation of these techniques fall beyond the scope of this paper and are left for future research.

## 5.6 Limitations of the evaluation

The primary purpose of the evaluation was to assess and validate, with users, a multisensory hybrid TV consumption experience based on the proposed approach, and to corroborate how multisensory effects enhance consumers' engagement, immersion, and QoE during the experience. This objective has been achieved, although the conducted study has some limitations, which are discussed in this section.

Regarding the evaluation settings, an ad-hoc prototype of a mulsemedia scenario [10] available in the authors' research laboratory was used. Although an attempt was made to recreate as realistic a mulsemedia scenario as possible by closing off the sides, front, and rear with opaque partition panels, the environment was not that of a home as such. Moreover, the number of sensory effect types evaluated was limited by the generation devices currently available in that scenario. Although no participants commented on this, the generation devices could not be properly 'camouflaged' in the scenario. Although the laboratory lights were turned off during the assessment sessions, their presence might have been somewhat intrusive.

Another limitation of the assessment was its reliance on a specific video clip that included only 6 content genres. The video clip content and its associated SEM metadata were subject to the limitations of the available mulsemedia platform. The authors wanted to test some of the available sensory effects in different TV genres in a short multisensory hybrid TV assessment session without extending it too much. The transitions between the different parts of the video clip (see Table 3) were designed to be as smooth as possible, using short video animations or static images to introduce the next part. While this format served the experimental purpose, it might lack the narrative coherence of professionally prepared TV content. The authors included as many sensory effects as possible, with minimal overlap, considering the available effects on the used mulsemedia platform and avoiding overloading the user with too many effects. Although more than 90% of participants agreed that the sensory effects were consistent with the AV content and well distributed and balanced throughout the TV video sequence, a few disagreed. Due to the short duration of the parts, some of them might seem somewhat overloaded with too many sensory effects, as the effects often overlapped or were very close together. A longer video with more genres and with the sensory effects more clearly separated in time, avoiding too much overlap, could have been more representative of a real multisensory hybrid TV session, but the evaluation sessions would have been too long. Although the overlap of effects in some parts of the employed video clip was not a problem during validation, as it did not annoy users or prompt complaints, it could have affected their experience and responses.

A few users reported minor discomfort or fatigue (especially among participants under 35), mainly due to the overly intense effects in certain parts of the video clip. Interestingly, a subset of this group of users reported that, despite feeling fatigue, they still enjoyed the overall mulsemedia experience. This apparent contradiction may be attributed to the dynamic nature of the content: while segments like the 'Mascletà' required a high concentration of intense effects (vibration and smoke scent), other sections, such as the sports news or the documentary, were significantly more relaxed.

The study also faced technical limitations regarding hardware and software. The mulsemedia experience, while comprehensive, was limited to currently available effects, omitting other possible effects (e.g., water spray or fog). Additionally, the effect generation devices were visible (see Figure 18) or even wearable (the haptic vest), and some produced operational noise. While this physical presence could be intrusive, no specific complaints were received from participants.

Regarding the participants, although the sample of 106 individuals is more than appropriate for this kind of study and achieved a perfect gender balance (50% female and 50% male), it was heavily skewed towards a young demographic, specifically the 16-25 age range (60.4% of the participants). Furthermore, because recruitment took place within a university environment, the profile was predominantly composed of university-educated participants (55.66%) and participants in active study or employment. Most possessed a high level of education and familiarity with technology. Had a different set of participants been used, with a much greater diversity in terms of age, employment status, education, level of familiarity with technology, or previous experience in mulsemedia environments, the results could be a bit different.

# 6 Conclusion and future work

In this paper, a fully MPEG-V- and HbbTV-compliant end-to-end solution for providing multi-sensorial hybrid (broadcast/broadband) TV content consumption experiences has been presented. A prototype following it has been implemented, replicating a real mulsemedia hybrid TV service, which has been subjectively validated and evaluated by 106 participants. Apart from valuable insights from them, highly satisfactory results have been obtained for QoE enhancement, synchronization of sensory effects, overall enjoyment, realism, and immersion, regardless of gender, age, or educational background. Sensory effects provided an enjoyable, immersive, and more realistic TV consumption experience for most participants. The results have also revealed interesting demographic patterns, such as participants with higher academic levels reporting lower satisfaction scores, distinct gender-based preferences regarding sensory acceptance, and potential barriers to commercial adoption, where realism and immersion do not translate into an equivalent willingness to pay for the enhanced services.

The authors would like to emphasize that the robustness of the multisensory hybrid TV experiences will depend on the long-term reliability of the employed mulsemedia platform at home. Reliable effect generation and microcontroller devices, a gateway device to connect the TV to them, all perfectly orchestrated and considering the user's preferences, will be needed at home. Modularity will be a key feature of the mulsemedia platform at home, enabling the addition of new effects and updates for next-generation devices. As far as the authors know, there are no commercially available mulsemedia platforms (like the one described) for purchase and home installation, but they expect such platforms to appear very soon.

As future work, the mulsemedia testbed will be extended to include higher-quality sensory effect generation devices, and additional subjective tests with content from more genres and additional sensory effects (e.g., fog or water-spray effects) will be conducted. The influence of overlap in sensory effects on users' QoE in multisensory hybrid TV services will also be analyzed. Participants will be better distributed across age ranges and user profiles (different study levels, technical and non-technical, etc.), and the results will be analyzed by gender, age range, and user profile in greater detail. The application of AI- based sensory effect adaptation or content-aware adaptation techniques to mitigate users' discomfort or distraction will also be studied.

# Acknowledgement

This publication was supported, in part, by the following grants with refs.: PID2021-126645OB-I00, funded by MICIU/AEI/10.13039/501100011033/ and by 'ERDF A way of making Europe'; and CIAICO/2022/025, funded by the Conselleria de Innovación, Universidades, Ciencia y Sociedad Digital of Generalitat Valenciana (DOGV 8919/05.10.2020). The authors would like to thank À Punt Media for providing some of the video content included in several parts of the broadcast video, and for granting permission to use it.

# APENDIX I

In this Appendix, a table listing the sensory effects included in the SEM file associated with the video clip used in the assessment, along with their properties, is provided.

*Table I.1. List of the sensory effects included in the SEM file associated with the prepared video clip*

| Nº | Effect Type | Position | Start (mm:ss.0) | End (mm:ss.0) | Intensity |
|---|---|---|---|---|---|
| 1 | Light | Center / Middle / Front | 00:00.0 | 17:40.0 | - |
| 2 | Scent (gunpowder) | Center-left / Middle / Front | 00:38.8 | 00:39.8 | 10.0 ml/h |
| 3 | Vibration | Center / Middle / Midway | 00:41.2 | 00:41.6 | 15.0 Hz |
| 4 | Vibration | Center / Bottom / Midway | 00:41.2 | 00:41.6 | 15.0 Hz |
| 5 | Vibration | Center / Middle / Midway | 00:42.2 | 00:44.4 | 15.0 Hz |
| 6 | Vibration | Center / Bottom / Midway | 00:42.2 | 00:44.4 | 15.0 Hz |
| 7 | Vibration | Center / Middle / Midway | 00:45.6 | 00:46.0 | 15.0 Hz |
| 8 | Vibration | Center / Bottom / Midway | 00:45.6 | 00:46.0 | 15.0 Hz |
| 9 | Vibration | Center / Middle / Midway | 00:47.8 | 00:48.1 | 15.0 Hz |
| 10 | Vibration | Center / Bottom / Midway | 00:47.8 | 00:48.1 | 15.0 Hz |
| 11 | Vibration | Center / Middle / Midway | 00:48.3 | 00:48.7 | 15.0 Hz |
| 12 | Vibration | Center / Bottom / Midway | 00:48.3 | 00:48.7 | 15.0 Hz |
| 13 | Vibration | Center / Middle / Midway | 00:51.8 | 00:52.1 | 15.0 Hz |
| 14 | Vibration | Center / Bottom / Midway | 00:51.8 | 00:52.1 | 15.0 Hz |
| 15 | Vibration | Center / Middle / Midway | 00:52.3 | 00:52.7 | 15.0 Hz |
| 16 | Vibration | Center / Bottom / Midway | 00:52.3 | 00:52.7 | 15.0 Hz |
| 17 | Vibration | Center / Middle / Midway | 00:55.9 | 00:56.2 | 15.0 Hz |
| 18 | Vibration | Center / Bottom / Midway | 00:55.9 | 00:56.2 | 15.0 Hz |
| 19 | Vibration | Center / Middle / Midway | 00:56.4 | 00:56.8 | 15.0 Hz |
| 20 | Vibration | Center / Bottom / Midway | 00:56.4 | 00:56.8 | 15.0 Hz |
| 21 | Vibration | Center / Middle / Midway | 00:59.9 | 01:00.2 | 15.0 Hz |
| 22 | Vibration | Center / Bottom / Midway | 00:59.9 | 01:00.2 | 15.0 Hz |
| 23 | Vibration | Center / Middle / Midway | 01:00.4 | 01:00.8 | 15.0 Hz |
| 24 | Vibration | Center / Bottom / Midway | 01:00.4 | 01:00.8 | 15.0 Hz |
| 25 | Vibration | Center / Middle / Midway | 01:04.1 | 01:04.4 | 15.0 Hz |
| 26 | Vibration | Center / Bottom / Midway | 01:04.1 | 01:04.4 | 15.0 Hz |
| 27 | Vibration | Center / Middle / Midway | 01:04.6 | 01:04.9 | 15.0 Hz |
| 28 | Vibration | Center / Bottom / Midway | 01:04.6 | 01:04.9 | 15.0 Hz |
| 29 | Vibration | Center / Middle / Midway | 01:07.9 | 01:08.2 | 15.0 Hz |
| 30 | Vibration | Center / Bottom / Midway | 01:07.9 | 01:08.2 | 15.0 Hz |
| 31 | Vibration | Center / Middle / Midway | 01:08.4 | 01:08.8 | 15.0 Hz |
| 32 | Vibration | Center / Bottom / Midway | 01:08.4 | 01:08.8 | 15.0 Hz |
| 33 | Vibration | Center / Middle / Midway | 01:11.9 | 01:12.3 | 15.0 Hz |
| 34 | Vibration | Center / Bottom / Midway | 01:11.9 | 01:12.3 | 15.0 Hz |
| 35 | Vibration | Center / Middle / Midway | 01:12.5 | 01:12.8 | 15.0 Hz |
| 36 | Vibration | Center / Bottom / Midway | 01:12.5 | 01:12.8 | 15.0 Hz |
| 37 | Vibration | Center / Middle / Midway | 01:15.8 | 01:16.1 | 15.0 Hz |
| 38 | Vibration | Center / Bottom / Midway | 01:15.8 | 01:16.1 | 15.0 Hz |
| 39 | Vibration | Center / Middle / Midway | 01:16.3 | 01:16.7 | 15.0 Hz |
| 40 | Vibration | Center / Bottom / Midway | 01:16.3 | 01:16.7 | 15.0 Hz |
| 41 | Vibration | Center / Middle / Midway | 01:19.8 | 01:20.1 | 15.0 Hz |
| 42 | Vibration | Center / Bottom / Midway | 01:19.8 | 01:20.1 | 15.0 Hz |
| 43 | Vibration | Center / Middle / Midway | 01:20.3 | 01:20.7 | 15.0 Hz |
| 44 | Vibration | Center / Bottom / Midway | 01:20.3 | 01:20.7 | 15.0 Hz |
| 45 | Vibration | Center / Middle / Midway | 01:24.1 | 01:25.8 | 15.0 Hz |
| 46 | Vibration | Center / Bottom / Midway | 01:24.1 | 01:25.8 | 15.0 Hz |
| 47 | Vibration | Center / Middle / Midway | 01:27.2 | 01:27.6 | 15.0 Hz |
| 48 | Vibration | Center / Bottom / Midway | 01:27.2 | 01:27.6 | 15.0 Hz |
| 49 | Vibration | Center / Middle / Midway | 01:28.0 | 01:28.3 | 15.0 Hz |
| 50 | Vibration | Center / Bottom / Midway | 01:28.0 | 01:28.3 | 15.0 Hz |
| 51 | Vibration | Center / Middle / Midway | 01:28.6 | 01:28.9 | 15.0 Hz |
| 52 | Vibration | Center / Bottom / Midway | 01:28.6 | 01:28.9 | 15.0 Hz |
| 53 | Vibration | Center / Middle / Midway | 01:31.6 | 01:31.9 | 15.0 Hz |
| 54 | Vibration | Center / Bottom / Midway | 01:31.6 | 01:31.9 | 15.0 Hz |
| 55 | Vibration | Center / Middle / Midway | 01:32.1 | 01:32.4 | 15.0 Hz |
| 56 | Vibration | Center / Bottom / Midway | 01:32.1 | 01:32.4 | 15.0 Hz |
| 57 | Vibration | Center / Middle / Midway | 01:34.9 | 01:35.3 | 15.0 Hz |
| 58 | Vibration | Center / Bottom / Midway | 01:34.9 | 01:35.3 | 15.0 Hz |
| 59 | Vibration | Center / Middle / Midway | 01:38.1 | 01:38.4 | 15.0 Hz |
| 60 | Vibration | Center / Bottom / Midway | 01:38.1 | 01:38.4 | 15.0 Hz |
| 61 | Vibration | Center / Middle / Midway | 01:38.7 | 01:39.0 | 15.0 Hz |
| 62 | Vibration | Center / Bottom / Midway | 01:38.7 | 01:39.0 | 15.0 Hz |
| 63 | Vibration | Center / Middle / Midway | 01:41.3 | 01:41.7 | 15.0 Hz |
| 64 | Vibration | Center / Bottom / Midway | 01:41.3 | 01:41.7 | 15.0 Hz |
| 65 | Vibration | Center / Middle / Midway | 01:41.9 | 01:42.2 | 15.0 Hz |
| 66 | Vibration | Center / Bottom / Midway | 01:41.9 | 01:42.2 | 15.0 Hz |
| 67 | Vibration | Center / Middle / Midway | 01:42.4 | 02:20.6 | 20.0 Hz |
| 68 | Vibration | Center / Bottom / Midway | 01:42.4 | 02:20.6 | 20.0 Hz |
| 69 | Vibration | Center / Middle / Midway | 02:20.9 | 02:21.6 | 20.0 Hz |
| 70 | Vibration | Center / Bottom / Midway | 02:20.9 | 02:21.6 | 20.0 Hz |
| 71 | Wind | Left / Top / Front | 02:47.8 | 02:54.0 | 2.0 Bft |
| 72 | Wind | Right / Top / Front | 02:47.8 | 02:54.0 | 2.0 Bft |
| 73 | Scent (fuel) | Center-left / Middle / Front | 02:47.8 | 02:48.8 | 10.0 ml/h |
| 74 | Wind | Left / Top / Front | 02:56.0 | 03:01.0 | 2.0 Bft |
| 75 | Wind | Right / Top / Front | 02:56.0 | 03:01.0 | 2.0 Bft |
| 76 | Wind | Left / Top / Front | 03:21.1 | 03:34.0 | 2.0 Bft |
| 77 | Wind | Left / Top / Back | 03:21.1 | 03:34.0 | 2.0 Bft |
| 78 | Scent (wet land) | Center-left / Middle / Front | 03:21.1 | 03:34.0 | 10.0 ml/h |
| 79 | Wind | Left / Top / Front | 03:39.1 | 03:45.6 | 2.0 Bft |
| 80 | Wind | Left / Top / Back | 03:39.1 | 03:45.6 | 2.0 Bft |
| 81 | Wind | Left / Top / Front | 03:50.1 | 03:54.0 | 2.0 Bft |
| 82 | Wind | Right / Top / Front | 03:50.1 | 03:54.0 | 2.0 Bft |
| 83 | Scent (cut grass) | Center-right / Middle / Front | 04:07.0 | 04:32.2 | 10.0 ml/h |
| 84 | Scent (cut grass) | Center-right / Middle / Front | 04:50.0 | 05:10.0 | 10.0 ml/h |
| 85 | Wind | Left / Top / Front | 06:14.3 | 06:15.0 | 8.0 Bft |
| 86 | Wind | Right / Top / Front | 06:14.3 | 06:15.0 | 8.0 Bft |
| 87 | Scent (burnt wheel) | Center-left / Middle / Front | 06:14.3 | 06:19.0 | 12.0 ml/h |
| 88 | Vibration | Center / Middle / Midway | 06:14.3 | 06:15.2 | 5.0 Hz |
| 89 | Vibration | Center / Bottom / Midway | 06:14.3 | 06:15.2 | 5.0 Hz |
| 90 | Vibration | Center / Middle / Midway | 06:17.3 | 06:18.1 | 5.0 Hz |
| 91 | Vibration | Center / Bottom / Midway | 06:17.3 | 06:18.1 | 5.0 Hz |
| 92 | Wind | Left / Top / Front | 06:17.2 | 06:17.8 | 8.0 Bft |
| 93 | Wind | Right / Top / Front | 06:17.2 | 06:17.8 | 8.0 Bft |
| 94 | Vibration | Center / Middle / Midway | 06:37.8 | 06:41.4 | 10.0 Hz |
| 95 | Vibration | Center / Bottom / Midway | 06:37.8 | 06:41.4 | 10.0 Hz |
| 96 | Scent (diesel) | Center-left / Middle / Front | 06:37.8 | 06:39.3 | 10.0 ml/h |
| 97 | Wind | Left / Top / Front | 06:39.2 | 06:46.2 | 8.0 Bft |
| 98 | Wind | Right / Top / Front | 06:39.2 | 06:46.2 | 8.0 Bft |
| 99 | Vibration | Center / Middle / Midway | 06:42.0 | 06:46.2 | 15.0 Hz |
| 100 | Vibration | Center / Bottom / Midway | 06:42.0 | 06:46.2 | 15.0 Hz |
| 101 | Wind | Left / Top / Front | 06:48.6 | 06:53.1 | 8.0 Bft |
| 102 | Wind | Right / Top / Front | 06:48.6 | 06:53.1 | 8.0 Bft |
| 103 | Vibration | Center / Middle / Midway | 06:48.6 | 06:53.1 | 15.0 Hz |
| 104 | Vibration | Center / Bottom / Midway | 06:48.6 | 06:53.1 | 15.0 Hz |
| 105 | Vibration | Center / Middle / Midway | 06:53.1 | 06:53.1 | 20.0 Hz |
| 106 | Vibration | Center / Bottom / Midway | 06:53.1 | 06:53.1 | 20.0 Hz |
| 107 | Vibration | Center / Middle / Midway | 06:57.4 | 06:57.8 | 20.0 Hz |
| 108 | Vibration | Center / Bottom / Midway | 06:57.4 | 06:57.8 | 20.0 Hz |
| 109 | Temperature | Right / Middle / Midway | 07:02.0 | 07:16.0 | 28.0 °C |
| 110 | Temperature | Left / Middle / Midway | 07:02.0 | 07:16.0 | 28.0 °C |
| 111 | Scent (chocolate) | Center-right / Middle / Front | 07:02.0 | 07:16.0 | 10.0 ml/h |
| 112 | Scent (orange) | Center-right / Middle / Front | 07:41.0 | 07:54.0 | 10.0 ml/h |
| 113 | Scent (pineapple) | Center-right / Middle / Front | 07:56.2 | 08:10.0 | 10.0 ml/h |
| 114 | Scent (banana) | Center-right / Middle / Front | 08:13.0 | 08:24.0 | 10.0 ml/h |
| 115 | Temperature | Right / Middle / Midway | 08:38.4 | 09:03.0 | 28.0 °C |
| 116 | Temperature | Left / Middle / Midway | 08:38.4 | 09:03.0 | 28.0 °C |
| 117 | Light (goldenrod) | Center / Top / Back | 08:38.4 | 09:03.0 | 200.0 lux |
| 118 | Scent (coffee chocolate) | Center-right / Middle / Front | 08:40.0 | 09:03.0 | 10.0 ml/h |
| 119 | Scent (bread) | Center-left / Middle / Front | 09:05.0 | 09:23.0 | 10.0 ml/h |
| 120 | Wind | Left / Top / Front | 09:41.0 | 10:20.0 | 2.0 Bft |
| 121 | Wind | Right / Top / Front | 09:41.0 | 10:20.0 | 2.0 Bft |
| 122 | Scent (wood) | Center-left / Middle / Front | 09:41.0 | 10:20.0 | 10.0 ml/h |
| 123 | Scent (pine tree) | Center-left / Middle / Front | 10:25.0 | 10:40.0 | 10.0 ml/h |
| 124 | Scent (lavender) | Center-left / Middle / Front | 10:41.0 | 11:00.0 | 10.0 ml/h |
| 125 | Scent (rosemary) | Center-right / Middle / Front | 11:05.0 | 11:30.0 | 10.0 ml/h |
| 126 | Temperature | Right / Middle / Midway | 11:32.4 | 11:43.0 | 28.0 °C |
| 127 | Temperature | Left / Middle / Midway | 11:32.4 | 11:43.0 | 28.0 °C |
| 128 | Scent (popcorn) | Center-right / Middle / Front | 11:33.2 | 11:43.0 | 10.0 ml/h |
| 129 | Light (amber) | Center / Top / Back | 11:35.0 | 11:43.0 | 110.0 lux |
| 130 | Light (green (color wheel) (x11 green)) | Center / Top / Back | 11:44.0 | 11:48.9 | 110.0 lux |
| 131 | Light (red) | Center / Top / Back | 11:49.8 | 11:50.3 | 110.0 lux |
| 132 | Light (blue) | Center / Top / Back | 11:51.2 | 11:51.7 | 110.0 lux |
| 133 | Light (white) | Center / Top / Back | 11:51.8 | 11:52.9 | 110.0 lux |
| 134 | Flash (white) | Center / Top / Back | 11:54.4 | 11:55.0 | 110.0 lux, 10 Hz |
| 135 | Wind | Left / Top / Front | 11:59.0 | 12:02.8 | 5.0 Bft |
| 136 | Wind | Right / Top / Front | 11:59.0 | 12:02.8 | 5.0 Bft |
| 137 | Temperature | Right / Middle / Midway | 11:59.0 | 12:28.0 | 28.0 °C |
| 138 | Temperature | Left / Middle / Midway | 11:59.0 | 12:28.0 | 28.0 °C |
| 139 | Vibration | Center / Middle / Midway | 12:03.5 | 12:04.3 | 20.0 Hz |
| 140 | Vibration | Center / Bottom / Midway | 12:03.5 | 12:04.3 | 20.0 Hz |
| 141 | Flash (white) | Center / Top / Back | 12:03.5 | 12:04.3 | 110.0 lux, 10 Hz |
| 142 | Vibration | Center / Middle / Midway | 12:08.0 | 12:09.0 | 30.0 Hz |
| 143 | Vibration | Center / Bottom / Midway | 12:08.0 | 12:09.0 | 30.0 Hz |
| 144 | Flash (white) | Center / Top / Back | 12:08.0 | 12:09.0 | 110.0 lux, 10 Hz |
| 145 | Vibration | Center / Middle / Midway | 12:12.5 | 12:13.4 | 30.0 Hz |
| 146 | Vibration | Center / Bottom / Midway | 12:12.5 | 12:13.4 | 30.0 Hz |
| 147 | Wind | Right / Top / Front | 12:20.4 | 12:21.3 | 3.0 Bft |
| 148 | Vibration | Center / Middle / Midway | 12:18.0 | 12:21.3 | 30.0 Hz |
| 149 | Vibration | Center / Bottom / Midway | 12:18.0 | 12:21.3 | 30.0 Hz |
| 150 | Vibration | Center / Middle / Midway | 12:23.0 | 12:23.4 | 20.0 Hz |
| 151 | Vibration | Center / Bottom / Midway | 12:23.0 | 12:23.4 | 20.0 Hz |
| 152 | Vibration | Center / Middle / Midway | 12:25.3 | 12:26.0 | 30.0 Hz |
| 153 | Vibration | Center / Bottom / Midway | 12:25.3 | 12:26.0 | 30.0 Hz |
| 154 | Vibration | Center / Middle / Midway | 12:27.0 | 12:28.0 | 30.0 Hz |
| 155 | Vibration | Center / Bottom / Midway | 12:27.0 | 12:28.0 | 30.0 Hz |
| 156 | Vibration | Center / Middle / Midway | 12:32.1 | 12:37.1 | 20.0 Hz |
| 157 | Vibration | Center / Bottom / Midway | 12:32.1 | 12:37.1 | 20.0 Hz |
| 158 | Scent (fuel) | Center-left / Middle / Front | 12:35.0 | 12:36.0 | 12.0 ml/h |
| 159 | Temperature | Right / Middle / Front | 12:41.1 | 12:57.0 | 18.0 °C |
| 160 | Temperature | Left / Middle / Front | 12:41.1 | 12:57.0 | 18.0 °C |
| 161 | Wind | Left / Top / Front | 12:46.0 | 12:47.7 | 10.0 Bft |
| 162 | Wind | Right / Top / Front | 12:46.0 | 12:47.7 | 10.0 Bft |
| 163 | Wind | Left / Top / Front | 12:51.0 | 12:53.0 | 10.0 Bft |
| 164 | Wind | Right / Top / Front | 12:51.0 | 12:53.0 | 10.0 Bft |
| 165 | Wind | Left / Top / Front | 12:55.8 | 12:57.6 | 10.0 Bft |
| 166 | Wind | Right / Top / Front | 12:55.8 | 12:57.6 | 10.0 Bft |
| 167 | Temperature | Right / Middle / Front | 13:16.7 | 15:03.0 | 18.0 °C |
| 168 | Temperature | Left / Middle / Front | 13:16.7 | 15:03.0 | 18.0 °C |
| 169 | Wind | Left / Top / Front | 13:45.0 | 14:07.0 | 8.0 Bft |
| 170 | Wind | Right / Top / Front | 13:45.0 | 14:07.0 | 8.0 Bft |
| 171 | Vibration | Center / Middle / Midway | 14:01.1 | 14:02.2 | 30.0 Hz |
| 172 | Vibration | Center / Bottom / Midway | 14:01.1 | 14:02.2 | 30.0 Hz |
| 173 | Wind | Left / Top / Front | 14:11.0 | 14:30.3 | 6.0 Bft |
| 174 | Wind | Right / Top / Front | 14:11.0 | 14:30.3 | 6.0 Bft |
| 175 | Wind | Left / Top / Front | 14:34.4 | 14:58.0 | 10.0 Bft |
| 176 | Wind | Right / Top / Front | 14:34.4 | 14:58.0 | 10.0 Bft |
| 177 | Vibration | Center / Middle / Midway | 14:34.9 | 14:35.6 | 20.0 Hz |
| 178 | Vibration | Center / Bottom / Midway | 14:34.9 | 14:35.6 | 20.0 Hz |
| 179 | Vibration | Center / Middle / Midway | 14:37.0 | 14:40.0 | 10.0 Hz |
| 180 | Vibration | Center / Bottom / Midway | 14:37.0 | 14:40.0 | 10.0 Hz |
| 181 | Vibration | Center / Middle / Midway | 14:47.6 | 14:48.3 | 20.0 Hz |
| 182 | Vibration | Center / Bottom / Midway | 14:47.6 | 14:48.3 | 20.0 Hz |
| 183 | Vibration | Center / Middle / Midway | 14:52.6 | 14:59.0 | 10.0 Hz |
| 184 | Vibration | Center / Bottom / Midway | 14:52.6 | 14:59.0 | 10.0 Hz |
| 185 | Wind | Left / Top / Front | 15:00.0 | 15:05.0 | 6.0 Bft |
| 186 | Wind | Right / Top / Front | 15:00.0 | 15:05.0 | 6.0 Bft |
| 187 | Temperature | Right / Middle / Midway | 15:10.0 | 15:15.0 | 28.0 °C |
| 188 | Temperature | Left / Middle / Midway | 15:10.0 | 15:15.0 | 28.0 °C |
| 189 | Wind | Left / Top / Front | 15:16.0 | 15:21.7 | 6.0 Bft |
| 190 | Wind | Right / Top / Front | 15:16.0 | 15:21.7 | 6.0 Bft |
| 191 | Light (blue) | Center / Top / Back | 15:22.9 | 15:28.7 | 110.0 lux |
| 192 | Scent (lavender) | Center-left / Middle / Front | 15:30.0 | 15:40.0 | 10.0 ml/h |
| 193 | Wind | Left / Top / Front | 15:43.0 | 15:48.9 | 6.0 Bft |
| 194 | Wind | Right / Top / Front | 15:43.0 | 15:48.9 | 6.0 Bft |
| 195 | Temperature | Right / Middle / Midway | 15:52.0 | 15:56.0 | 28.0 °C |
| 196 | Temperature | Left / Middle / Midway | 15:52.0 | 15:56.0 | 28.0 °C |
| 197 | Scent (burnt wood) | Center-right / Middle / Front | 15:52.0 | 15:56.0 | 10.0 ml/h |
| 198 | Temperature | Right / Middle / Midway | 16:27.8 | 16:32.0 | 28.0 °C |
| 199 | Temperature | Left / Middle / Midway | 16:27.8 | 16:32.0 | 28.0 °C |
| 200 | Scent (burnt wood) | Center-right / Middle / Front | 16:27.8 | 16:32.0 | 12.0 ml/h |
| 201 | Temperature | Right / Middle / Midway | 16:43.6 | 16:45.2 | 28.0 °C |
| 202 | Temperature | Left / Middle / Midway | 16:43.6 | 16:45.2 | 28.0 °C |
| 203 | Vibration | Center / Middle / Midway | 16:47.6 | 16:48.2 | 20.0 Hz |
| 204 | Vibration | Center / Bottom / Midway | 16:47.6 | 16:48.2 | 20.0 Hz |
| 205 | Wind | Left / Top / Front | 17:00.0 | 17:07.7 | 4.0 Bft |
| 206 | Wind | Right / Top / Front | 17:00.0 | 17:07.7 | 4.0 Bft |
| 207 | Vibration | Center / Middle / Midway | 17:00.0 | 17:02.2 | 20.0 Hz |
| 208 | Vibration | Center / Bottom / Midway | 17:00.0 | 17:02.2 | 20.0 Hz |
| 209 | Vibration | Center / Middle / Midway | 17:07.3 | 17:07.8 | 20.0 Hz |
| 210 | Vibration | Center / Bottom / Midway | 17:07.3 | 17:07.8 | 20.0 Hz |
| 211 | Temperature | Right / Middle / Midway | 17:08.0 | 17:23.0 | 28.0 °C |
| 212 | Temperature | Left / Middle / Midway | 17:08.0 | 17:23.0 | 28.0 °C |
| 213 | Scent (burnt wood) | Center-right / Middle / Front | 17:08.0 | 17:23.0 | 12.0 ml/h |
| 214 | Vibration | Center / Middle / Midway | 17:14.6 | 17:15.1 | 20.0 Hz |
| 215 | Vibration | Center / Bottom / Midway | 17:14.6 | 17:15.1 | 20.0 Hz |
| 216 | Vibration | Center / Middle / Midway | 17:16.4 | 17:24.0 | 10.0 Hz |
| 217 | Vibration | Center / Bottom / Midway | 17:16.4 | 17:24.0 | 20.0 Hz |
| 218 | Temperature | Right / Middle / Midway | 17:30.0 | 17:35.0 | 28.0 °C |
| 219 | Temperature | Left / Middle / Midway | 17:30.0 | 17:35.0 | 28.0 °C |
| 220 | Light (orange-red) | Center / Top / Back | 17:30.0 | 17:35.0 | 100.0 lux |